\documentclass[12pt,noinfoline]{imsart}

\usepackage{amsmath,amssymb,amsthm,amscd,amsfonts,bm,amsbsy}
\usepackage{graphicx}%
\usepackage{fancyhdr}
\usepackage{geometry}
\usepackage{url}
\usepackage{color}
\usepackage[colorlinks,citecolor=blue,urlcolor=blue]{hyperref}

\usepackage{natbib}
\usepackage{wrapfig,verbatim}
\usepackage{cutwin}
\usepackage{blkarray}
\usepackage{enumitem}
\usepackage{tikz-cd}
\usepackage{tikz}
\usepackage{mathtools}
\usepackage[ruled,vlined]{algorithm2e}
\usepackage{thmtools}
\usepackage{thm-restate}
\usepackage{dsfont}

\theoremstyle{plain} %\numberwithin{equation}{section}
\newtheorem{thm}{Theorem}[section]
\newtheorem{corollary}[thm]{Corollary}

\newtheorem{conj}{Conjecture}
\newtheorem{question}{Question}
\newtheorem{lemma}[thm]{Lemma}
\newtheorem{proposition}[thm]{Proposition}
\newtheorem{prop}[thm]{Proposition}

\theoremstyle{definition}
\newtheorem{defn}[thm]{Definition}

\newtheorem{remark}[thm]{Remark}
\newtheorem{rmk}[thm]{Remark}
\newtheorem{example}[thm]{Example}

\newcommand{\R}{\mathbb{R}}

\newcommand{\N}{\mathbb{N}}
\newcommand{\Z}{\mathbb{Z}}

\newcommand{\vphi}{\varphi}

\newcommand{\ep}{\epsilon}
\newcommand{\minus}{\backslash}

\DeclareMathOperator*{\argmax}{arg\,max}
\newcommand{\DC}{\mathrel{\ooalign{$D$\hfil\cr\kern.57em $C$}}}
\newcommand{\IDC}{I_{\scriptsize{\DC}_{\!n,z}}}
\newcommand{\fiberDC}{\mathcal{F}_{\scriptsize{\DC}_{\!n,z}}}
\newcommand{\sfiberDC}{\widetilde{\mathcal F}_{\scriptsize{\DC}_{\!n,z}}}

\definecolor{Rcolor}{RGB}{214, 39, 40}    % Red vertices
\definecolor{Gcolor}{RGB}{44, 160, 44}    % Green vertices
\definecolor{Bcolor}{rgb}{0.0, 0.5, 1.0}  % Blue vertices

\definecolor{forestgreen}{rgb}{0,.72,0} 
\definecolor{brickred}{rgb}{.72,0,0}
\definecolor{darkcerulean}{rgb}{0.03, 0.27, 0.49}

\begin{document}

\title{Irreducible Markov Chains on spaces of graphs with fixed degree-color sequences}

\author{\fnms{F\'elix Almendra-Hern\'andez}\thanksref{addr1}\corref{}\ead[label=e1]{falmendrahernandez@ucdavis.edu}},
\author{\fnms{Jes\'us A. De Loera}\thanksref{addr1}\ead[label=e2]{deloera@math.ucdavis.edu}},
\author{\fnms{Sonja Petrovi\'c}\thanksref{addr2}\ead[label=e3]{sonja.petrovic@illinoistech.edu}}

\thankstext{addr1}{Department of Mathematics, University of California, Davis}
\thankstext{addr2}{Department of Applied Mathematics, Illinois Institute of Technology. Corresponding author sonja.petrovic@illinoistech.edu}

% \author{F\'elix Almendra-Hernández, Jes\'us A. De Loera, Sonja Petrovi\'c}\thanks{FAH and JDL are at UC Davis. SP is at Illinois Institute of Technology and is partially supported by the Simons Foundation Collaboration Grant for Mathematicians \#854770 and DOE/SC award \#1010629.}
% \date{\today}
%\tableofcontents   

\begin{abstract} 
We study a colored generalization of the famous simple-switch Markov chain for sampling the set of graphs with a fixed degree sequence. Here we consider the space of graphs with colored vertices, in which we fix the degree sequence and another statistic arising from the vertex coloring, %also the number of edges between each color pair, 
 and prove that the set can be connected with simple color-preserving switches or moves.  These moves form a basis for defining an  irreducible Markov chain necessary for testing  statistical model fit to block-partitioned network data. 
Our methods further generalize well-known  algebraic results from the 1990s: namely,  that  the corresponding moves can be used to construct a regular triangulation for a generalization of the second hypersimplex.
On the other hand, in  contrast to the monochromatic case, we show that for \emph{simple} graphs,   the 1-norm of the moves necessary to connect the space increases with the number of colors. 

\end{abstract} 

\maketitle

\section{Introduction} 

A \emph{simple graph} is a pair $G=(V,E)$ with vertices  $V=[n]:=\{1,\ldots, n\}$  and edge set $E$. 
A \emph{multigraph} is a pair $G=(V,E)$  with vertex set $V=[n]$ in which $E$ is a multiset, so that there can be more than one edge between any two vertices in $V$.  
Notice that every simple graph can be viewed as a multigraph with at most one edge between nodes. 
In this paper, we do not allow self-loops, which are edges from a node to itself. 
If not explicitly specified, when we say `graph', we could be referring to either a simple or multigraph; each case will be clear from the context.  
A sequence $d\in \N^n$ is called \emph{graphical} if there exists a multigraph or a simple graph $G$, depending from the context, such that $d=d(G)$.
Graphical degree sequences have been extensively investigated by numerous authors across various fields and from diverse perspectives; see, for example, \cite{erdos1960,Peledetal89Polytope_deg_seq,Stanley91DegSeq,Palmer2000RandomTop,Cooperetal2003WebGraphs,Deza2018Opt_deg_seq,Berczietal2019Packing,DeedsEtAl2023Database,onn_degree_2023} and references therein. 

A particularly intriguing area of research centers around the random generation of graphs with a fixed degree sequence (see \cite{KTV1997degseq,Adbullah2010Covertime,Chatterjeeetal2011,Cloteaux2016Sequential,Greenhill22unified}). 
\cite{KTV1997degseq} proposed the use of the \emph{switch Markov chain} to uniformly generate simple graphs with a fixed degree sequence. Informally, a \emph{switch} is an operation that exchanges a pair of edges on four vertices with another pair of edges between the same four vertices in a way that the degree sequence of the graph is preserved.  As mentioned in \cite{Greenhill22unified} ``\textit{the switch Markov chain can be thought of as the Markov chain of smallest possible modifications}''.

It has long been established that the set of all simple graphs with fixed degree sequences can be connected through switches. Notably, \cite{havel1955remark} and \cite{hakimi1962realizability} leveraged this insight to provide a constructive solution to the \emph{graph realization problem}. This solution is commonly referred to as the \emph{Havel-Hakimi algorithm}. Alternatively, to test whether a sequence is realizable by a simple graph, one can employ the well-known \emph{Erd\H{o}s-Gallai theorem} (\cite{erdos1960}).

When examining the random generation problem from an algebraic statistics perspective, it is useful to conceptualize the space of simple graphs with a fixed degree sequence $d\in\N^n$ as a set of 0/1 vectors $\gamma\in \{0,1\}^{\binom{n}{2}}$ satisfying the system of linear equations $D_n \gamma=d$ where $D_n$ is the incidence matrix of the complete graph $K_n$. In this sense, the degree sequence is a linear graph statistic. If we relax the 0/1 constraints to allow multiple edges, the task of constructing an irreducible Markov chain on the space of multigraphs with a fixed degree sequence translates into the task of sampling integer points inside the convex polytope $\{\gamma\in \N^{\binom{n}{2}}: D_n\gamma=d\}$. This problem can be approached using algebraic techniques, as we explain later. Here, $\gamma\in \N^{\binom{n}{2}}$ represents the \emph{vectorization} of the multigraph $G$ with $\gamma_{uv}$ edges between $u$ and $v$. In this manuscript, we delve into a colored generalization of the connectivity problem on spaces of graphs with a fixed degree sequence and a fixed graph statistic arising from a vertex coloring. We explore this connectivity problem using the theory of Markov bases (see Definition~\ref{defn:MB}).

\begin{defn}
For a positive integer $k$ and a \emph{$k$-coloring} $z:[n]\to [k]$, we define the \emph{color sequence} of a graph $G$ to be the vector $c(G, z):=(c(z, i,j): 1\leq i\leq j \leq k)$ with $c(z,i,j)$ being equal to the number of edges in $G$ joining colors $i$ and $j$. The entries of $c(G,z)$ are ordered lexicographically with respect to the pairs $(i,j)$ and when $z$ is clear from the context we simply write $c(i,j)$ and $c(G)$. The \emph{degree-color sequence} $(d(G);c(G))$ of $G$ with a given $k$-coloring $z$, is the concatenation of its degree and color sequences. For simplicity, we call $(d(G);c(G))$ the \emph{c-degree sequence} from now on. We say that a sequence $(d;c)\in \N^{n+\binom{k+1}{2}}$ is \emph{color-graphical} if there is a multigraph or a simple graph $G$, depending from the context, and a $k$-coloring of $V(G)$ such that $(d;c) = (d(G);c(G))$.
\end{defn}

\begin{figure}[h]
\centering
\begin{tikzpicture}[scale=1, every node/.style={minimum size=5mm, text centered}]
   % Draw the circular graph
   \foreach \i/\color in {1/Bcolor, 2/Bcolor, 3/Rcolor, 4/Rcolor, 5/Gcolor} {
       \node[draw, circle, preaction={fill=\color, fill opacity=0.1}, draw=\color, inner sep=0pt] (n\i) at ({(-\i)*360/5+72}:1.7) {{\i}};
   }
   
   % Draw the edges
    \draw (n1) edge (n2);
    \draw (n1) edge[bend left=15] (n3);
    \draw (n1) edge[bend left=-15] (n3);
    \draw (n1) edge (n5);
    \draw (n2) edge (n4);
    \draw (n2) edge[bend left=15] (n5);
    \draw (n2) edge[bend left=-15] (n5);
    \draw (n3) edge (n5);
    \draw (n4) edge (n5);
    \draw (n4) edge[bend left=20] (n5);
    \draw (n4) edge[bend left=-20] (n5);

\end{tikzpicture}
\caption{Graph $G$ with vertices 1 and 2 colored blue, 3 and 4 red, and 5 green.} 
\label{fig:example-graph}
\end{figure}
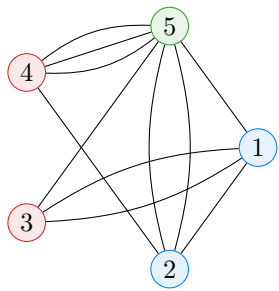

\begin{example}\label{example:graph_def}
For $n=5$ and $k=3$ let $\{\{1,2\},\{3,4\},\{5\}\}$ be the partition of $[5]$ induced by a $3$-coloring $z$ of $[5]$. \begin{tikzpicture}[every node/.style={minimum size=.3mm}] \node[draw, circle, draw=Bcolor, fill=Bcolor, fill opacity=0.2, inner sep=1.8pt] (x) at (0,0){};\end{tikzpicture}, \begin{tikzpicture}[every node/.style={minimum size=.3mm}] \node[draw, circle, draw=Rcolor, fill=Rcolor, fill opacity=0.2, inner sep=1.8pt] (x) at (0,0){};\end{tikzpicture} and \begin{tikzpicture}[every node/.style={minimum size=.3mm}] \node[draw, circle, draw=Gcolor, fill=Gcolor, fill opacity=0.2, inner sep=1.8pt] (x) at (0,0){};\end{tikzpicture} represent colors 1,2 and 3, respectively. The $c$-degree sequence of the graph $G$ illustrated in Figure~\ref{fig:example-graph} is the vector in $\N^{11}$ given by $(d(G); c(G))= (4, 4,3,4,7; 1,3,3,0,4,0)$.
\end{example}

The $c$-degree sequence is also a linear graph statistic, as we explain. For a $k$-coloring $z$ of $[n]$, we define the matrix $C_z$. Its rows are indexed by the $\binom{k+1}{2}$ pairs of colors (allowing for repetition), while its columns are indexed by the $\binom{n}{2}$ distinct pairs of vertices. Each column contains exactly one 1 in the row corresponding to that vertex pair's color pair, with the remaining entries of the column set to zero. For the \text{vectorization} $\gamma\in \N^{\binom{n}{2}}$ of a graph $G$, the color sequence of $G$ can be expressed as $c(G)=C_z\gamma$. Consequently, the $c$-degree sequence, $(d(G);c(G))$, can be written as $(D_n\gamma; C_z\gamma)=\DC_{\!n,z}\!\gamma$, where $\DC_{\!n,z}:=\begin{pmatrix}D_n \\ C_z\end{pmatrix}$. Following Definition~\ref{defn:MB}, this means that for a given $(d;c)\in \N^{n+\binom{k+1}{2}}$, the $(d;c)$-fiber $\fiberDC(d;c)=\{\gamma\in \N^{\binom{n}{2}}:\DC_{\!n,z}\!\gamma=(d;c)\}$ represents the set of multigraphs on $[n]$ with the fixed $c$-degree sequence $(d;c)$.

\begin{example}\label{example:matrix} Let $n=5, k=3$ and let $z$ be the $3$-coloring used in Example~\ref{example:graph_def}. The matrix $\DC_{\!n,z}$ is explicitly written below with a picture of $K_5$ on the left that helps visualize the encoding of the matrix. 

\vspace{.05in}
\begin{minipage}{0.3\textwidth}
    \centering
    \begin{tikzpicture}[scale=1, every node/.style={minimum size=5mm, text centered}]
       
       % Draw the circular graph
       \foreach \i/\color in {1/Bcolor, 2/Bcolor, 3/Rcolor, 4/Rcolor, 5/Gcolor} {
           \node[draw, circle, preaction={fill=\color, fill opacity=0.1}, draw=\color, inner sep=0pt] (n\i) at ({(-\i)*360/5+72}:1.75) {{\i}};
       }
       
       % Draw the edges
        \draw (n1) edge (n2);
        \draw (n1) edge (n3);
        \draw (n1) edge (n4);
        \draw (n1) edge (n5);
        \draw (n2) edge (n3);
        \draw (n2) edge (n4);
        \draw (n2) edge (n5);
        \draw (n3) edge (n4);
        \draw (n3) edge (n5);
        \draw (n4) edge (n5);

    \end{tikzpicture}
\end{minipage}%
\begin{minipage}{0.7\textwidth}
    \raggedright

    \definecolor{Rcolor}{RGB}{214, 39, 40}    % Red vertices
    \definecolor{Gcolor}{RGB}{44, 160, 44}    % Green vertices
    \definecolor{Bcolor}{rgb}{0.0, 0.5, 1.0}  % Blue vertices

    \[
    \DC_{5,z}=\begin{blockarray}{ccccccccccc}
    \textcolor{Bcolor}{\scriptstyle 1}\textcolor{Bcolor}{\scriptstyle 2} &  \textcolor{Bcolor}{\scriptstyle 1}\textcolor{Rcolor}{\scriptstyle 3} &  \textcolor{Bcolor}{\scriptstyle 1}\textcolor{Rcolor}{\scriptstyle 4}  &  \textcolor{Bcolor}{\scriptstyle 1} \textcolor{Gcolor}{\scriptstyle 5} & \textcolor{Bcolor}{\scriptstyle 2}\textcolor{Rcolor}{\scriptstyle 3} & \textcolor{Bcolor}{\scriptstyle 2}\textcolor{Rcolor}{\scriptstyle 4} & \textcolor{Bcolor}{\scriptstyle 2} \textcolor{Gcolor}{\scriptstyle 5} & \textcolor{Rcolor}{\scriptstyle 3}\textcolor{Rcolor}{\scriptstyle 4}  & \textcolor{Rcolor}{\scriptstyle 3} \textcolor{Gcolor}{\scriptstyle 5} & \textcolor{Rcolor}{\scriptstyle 4} \textcolor{Gcolor}{\scriptstyle 5}\\
    \begin{block}{(cccccccccc)c}
    1& 1& 1& 1& 0& 0& 0& 0& 0& 0& \textcolor{Bcolor}{\scriptstyle 1}
    \\ 
    1& 0& 0& 0& 1& 1& 1& 0& 0& 0& \textcolor{Bcolor}{\scriptstyle 2}\\ 
    0& 1& 0& 0& 1& 0& 0& 1& 1& 0& \textcolor{Rcolor}{\scriptstyle 3}\\ 
    0& 0& 1& 0& 0& 1& 0& 1& 0& 1& \textcolor{Rcolor}{\scriptstyle 4}\\ 
    0& 0& 0& 1& 0& 0& 1& 0& 1& 1& \textcolor{Gcolor}{\scriptstyle 5}\\\cline{1-10} 
    1& 0& 0& 0& 0& 0& 0& 0& 0& 0& \begin{tikzpicture}[every node/.style={minimum size=.1mm}] \node[draw, circle, draw=Bcolor, fill=Bcolor, fill opacity=0.2, inner sep=1.25pt] (x) at (0,0){}; \node[draw, circle, draw=Bcolor, fill=Bcolor, fill opacity=0.2, inner sep=1.25pt] (y) at (.3,0){}; \draw (x) edge (y); \end{tikzpicture}
    \\ 
    0& 1& 1& 0& 1& 1& 0& 0& 0& 0& \begin{tikzpicture}[every node/.style={minimum size=.1mm}] \node[draw, circle, draw=Bcolor, fill=Bcolor, fill opacity=0.2, inner sep=1.25pt] (x) at (0,0){}; \node[draw, circle, draw=Rcolor, fill=Rcolor, fill opacity=0.2, inner sep=1.25pt] (y) at (.3,0){}; \draw (x) edge (y); \end{tikzpicture}\\ 
    0& 0& 0& 1& 0& 0& 1& 0& 0& 0& \begin{tikzpicture}[every node/.style={minimum size=.1mm}] \node[draw, circle, draw=Bcolor, fill=Bcolor, fill opacity=0.2, inner sep=1.25pt] (x) at (0,0){}; \node[draw, circle, draw=Gcolor, fill=Gcolor, fill opacity=0.2, inner sep=1.25pt] (y) at (.3,0){}; \draw (x) edge (y); \end{tikzpicture}\\
    0& 0& 0& 0& 0& 0& 0& 1& 0& 0& \begin{tikzpicture}[every node/.style={minimum size=.1mm}] \node[draw, circle, draw=Rcolor, fill=Rcolor, fill opacity=0.2, inner sep=1.25pt] (x) at (0,0){}; \node[draw, circle, draw=Rcolor, fill=Rcolor, fill opacity=0.2, inner sep=1.25pt] (y) at (.3,0){}; \draw (x) edge (y); \end{tikzpicture}\\
    0& 0& 0& 0& 0& 0& 0& 0& 1& 1& \begin{tikzpicture}[every node/.style={minimum size=.1mm}] \node[draw, circle, draw=Rcolor, fill=Rcolor, fill opacity=0.2, inner sep=1.25pt] (x) at (0,0){}; \node[draw, circle, draw=Gcolor, fill=Gcolor, fill opacity=0.2, inner sep=1.25pt] (y) at (.3,0){}; \draw (x) edge (y); \end{tikzpicture}\\
    0& 0& 0& 0& 0& 0& 0& 0& 0& 0& \begin{tikzpicture}[every node/.style={minimum size=.1mm}] \node[draw, circle, draw=Gcolor, fill=Gcolor, fill opacity=0.2, inner sep=1.25pt] (x) at (0,0){}; \node[draw, circle, draw=Gcolor, fill=Gcolor, fill opacity=0.2, inner sep=1.25pt] (y) at (.3,0){}; \draw (x) edge (y); \end{tikzpicture}\\
    \end{block}
    \end{blockarray}
\]
\end{minipage}

\end{example}

The $c$-degree sequence arises as the sufficient statistic of a random network model in which each edge in the graph appears with a probability that depends on its endpoint vertices as well as their color. In statistics, colors represent blocks or communities, and the model is called the \emph{$\beta$-SBM}, where SBM stands for the stochastic blockmodel for random graphs, initially introduced in the work of \cite{holland1983stochastic}. Variants of the SBM, inspired by the Erd\H{o}s-R\'enyi model, were outlined by Fienberg, Meyer, and Wasserman in \cite{FienbergMeyerWasserman1985block} and it has been the blueprint model for  social, economic, protein, and neural network data  with a given block structure. Goodness of fit of the $\beta$-SBM  was studied in \cite{karwa2016exact}, which gave a dynamic algorithm for the test and discussed how to use Markov bases to extend the exact test to latent blocks version of the model. However, they left the determination of a Markov basis for its \emph{design matrix $\DC_{\!n,z}$} an open problem, stated as \cite[Conjecture 1]{karwa2016exact}. Here we address this question.

\begin{defn}\label{defn:MB} For an integer matrix $A\in \Z^{r\times c}$ and a vector $b\in \Z^{r}$, we call the set of integer points $\mathcal F_A(b):=\{x\in \N^c: Ax=b\}$ the $b$-\emph{fiber} of the matrix $A$. For any finite set of \emph{moves} $\mathcal B\subset \ker_\Z A$, we define $\mathcal F_A(b)_{\mathcal B}$ as the graph  whose vertices are the elements of $\mathcal F_A(b)$ and any two vertices $m,m'\in \mathcal F_A(b)$ form an edge if $m-m'\in \pm\mathcal B$. The set $\mathcal B$ is called a \emph{Markov basis} for $A$ if $\mathcal F_A(b)_{\mathcal B}$ is connected for every $b$. 
\end{defn}

Intuitively, a Markov basis is a collection of moves that guarantees connectivity for non-negative integer solutions across many different families of systems of linear equations.
Returning to our specific problem, if we let $\mathcal{M}_{n,z}:=\{\gamma\in \ker_\Z\!\DC_{\!n,z}: ||\gamma||_1=4\}$, we prove the following result.

\begin{restatable}{thm}{thmone}\label{thm:generators}
The set of quadratic moves $\mathcal M_{n,z}$ is a Markov basis for $\DC_{\!n,z}$. These are the moves in $\ker_\Z\!\DC_{\!n,z}$ of minimal 1-norm.  
\end{restatable}

Similar to the monochromatic case, this Markov basis is equivalent to the set of smallest possible modifications. In essence, any two multigraphs with a fixed $c$-degree sequence can be connected by applying a sequence of $c$-degree-preserving switches of 4 edges at a time. A natural follow-up question is whether the connectivity in the space of multigraphs with a fixed $c$-degree sequence, induced by the moves in Theorem~\ref{thm:generators}, is maintained when restricting to the space of simple graphs.

For any color-graphical $c$-degree sequence $(d;c)\in \N^{n+\binom{k+1}{2}}$ we define the \emph{simple $(d;c)$-fiber} of $\DC_{\!n,z}$ as the set $\sfiberDC(d;c):=\fiberDC(d;c)\cap \{0,1\}^{\binom{n}{2}}$ of all simple graphs with fixed $c$-degree sequence $(d;c)$. We say that a set of moves $\mathcal B\subset \ker_\Z\!\DC_{\!n,z}$ is a \emph{simple-graph Markov basis} for $\DC_{\!n,z}$ if for every color-graphical $c$-degree sequence $(d;c)\in \{0,1\}^{n+\binom{k+1}{2}}$, the induced graph $\sfiberDC(d;c)_{\mathcal B}$ is connected. Using this notation, we provide a negative answer to the posed question.

\begin{restatable}{thm}{thmtwo}\label{thm:no_universal_constant} For any constant $C$, there exists $n,k\in \Z_+$ and a $k$-coloring $z$ of $[n]$ such that any simple-graph Markov basis for $\DC_{\!n,z}$ has an element with 1-norm greater than $C$.  
\end{restatable}

In contrast to the behavior observed in the monochromatic case, the 1-norm size of the moves necessary to connect spaces of simple graphs with a fixed $c$-degree sequence increases as the number of colors $k$ used in the $k$-coloring $z$ varies. Motivated by this result, we present an open problem and a conjecture at the end of Section~\ref{section:simple_graphs}. It is important to notice that a simple-graph Markov basis might not be a Markov basis in the sense of Definition~\ref{defn:MB} as these two sets of moves connect different spaces of graphs.

\begin{figure}[h!]
    \includegraphics[scale=.72]{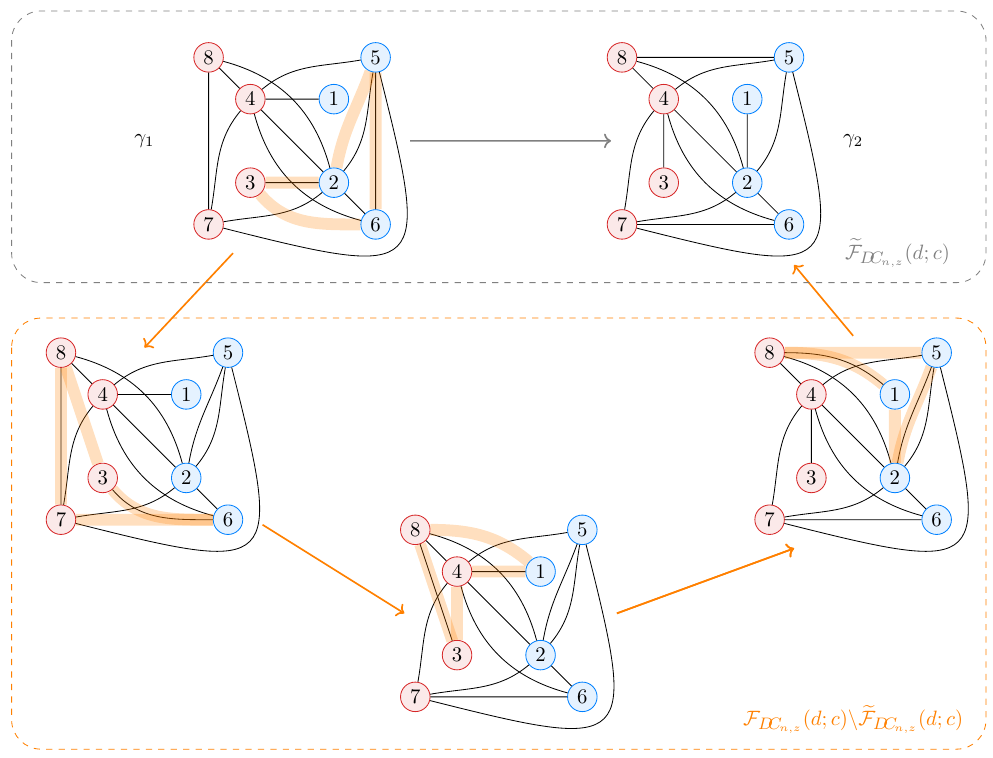}
\caption{Simple graphs $\gamma_1$ and $\gamma_2$ being connected with switches by leaving the simple-graph fiber. The switches used in each step are highlighted in orange.}
\label{fig:connect_indispensable_squares}
\end{figure}

\begin{example}\label{example:connecting_multigraphs} The simple graphs $\gamma_1, \gamma_2$ in Figure~\ref{fig:connect_indispensable_squares} represent the only two elements of the simple-graph fiber $\widetilde{\mathcal F}_{\scriptsize{\DC}_{\!8,z}}(d; c)$ where $z$ is a $2$-coloring that induces the partition $\{\{1,2,5,6\},\allowbreak \{3,4,7,8\}\}$, $d=(1,6,1,6,4,3,4,3)$ and $c=(3,8,3)$. The only move (up to sign) that connects this simple-graph fiber is $\gamma=\gamma_1-\gamma_2$, whose 1-norm is 8. We illustrate one way of connecting $\gamma_1$ to $\gamma_2$ using elements from $\mathcal M_{n,z}$ by stepping into $\mathcal{F}_{\scriptsize{\DC}_{\!n,z}}\minus\widetilde{\mathcal F}_{\scriptsize{\DC}_{\!n,z}}$. The orange highlighted edges in each graph explain what switch is performed to reach the next graph in the orange path.
\end{example} 

To explore more of the advantageous properties of switches preserving $c$-degree sequences, we delve into their algebraic analogues that were introduced by \cite{DS98}. By these analogues, we refer to the fact that every integer vector $\gamma\in \Z^{\binom{n}{2}}$ corresponds to a \emph{binomial} in the ring of multivariate polynomials $K[x_{uv}:1\leq u< v\leq n]$, given by $x^{\gamma^+}-x^{\gamma^-}$, where $\gamma^+=\max(\gamma, 0)$ and $\gamma^-=\max(-\gamma, 0)$ entrywise. We interpret the binomial $x^{\gamma^+}-x^{\gamma^-}$ as an encoding of the operation that exchanges the edges in $\gamma^+$ with edges in $\gamma^-$. Before explicitly stating our result involving these algebraic analogues, we provide a quick tool overview.

In general, for a polynomial ring $K[x_1, \ldots, x_s]$ and an ideal $I\subset K[x_1, \ldots, x_s]$, there is always a finite set of generators for $I$ with powerful algorithmic properties. These special sets of generators are known as \emph{Gr\"obner bases}, and they have played a major role in the development of computational commutative algebra and algebraic geometry. Given the prevalence of polynomial models in various scientific and engineering disciplines, Gr\"obner bases have found countless applications across numerous domains. One particularly interesting application for our purposes is presented in \cite{DS98}, where they describe Markov chain algorithms for sampling from conditional distributions of discrete exponential families. 

To formally define a Gr\"obner basis we first introduce the following notions. A \emph{monomial order} $\succ$ on $K[x_1, \ldots, x_s]$ is any relation $\succ$ on the set of \emph{monomials} $x^\alpha$, $\alpha\in \N^{s}$ satisfying the following conditions: $\succ$ is a total order, if $x^\alpha\succ x^{\theta}$ then $x^{\alpha+\gamma}\succ x^{\theta+\gamma}$ for every $\gamma\in \N^s$, and every nonempty set of monomials has a smallest element under $\succ$. The \emph{leading monomial} of a polynomial $f\in K[x_1, \ldots, x_s]$ with respect to the monomial order $\succ$ will be denoted by $\text{in}_\succ(f)$ and for any ideal $I\subset K[x_1, \ldots, x_s]$, $\text{in}_\succ(I):=\langle\{\text{in}_\succ(f):f \in I\}\rangle$ is the ideal generated by all the leading monomials of $I$. We say that a finite subset of polynomials $\mathcal G$ is a \emph{Gr\"obner basis} for $I$ with respect to the monomial order $\succ$ if $\text{in}_\succ(I)=\langle \text{in}_\succ(g): g\in \mathcal G\rangle$. This means that $\mathcal G$ is a G\"obner basis for $I$ with respect to $\succ$ if for any polynomial $f\in I$, $\text{in}_\succ(f)$ is divisible by $\text{in}_\succ(g)$ for some $g\in \mathcal G$. One particular property is that any Gr\"obner basis of $I$, generates the ideal $I$. For a more detailed introduction to the subject see \cite{whatisgroebner2005} and \cite{CLO}. 

Using an integer matrix $A=(a_{\upsilon,\nu})\in \Z^{r\times c}$, one can define a particularly special polynomial ideal known as \emph{toric ideal}, denoted by $I_A$ (see \cite[Chapter 4]{St}). This is the kernel of the semigroup algebra homomorphism
\begin{align}\label{eqn:toricmap}
\notag 
    \varphi: K[x_{1}, \ldots, x_c] &\rightarrow K[t_1^{\pm}, \ldots, t_r^{\pm}];\\
     x_{\nu} &\mapsto \prod_{\upsilon=1}^rt_\upsilon^{a_{\upsilon,\nu}}. 
\end{align}
In the algebraic statistics literature, the matrix $A$ defines  the statistical log-linear model $\mathcal M_A:=\{p=(p_1,\ldots, p_r)\in\R^r: \sum_{\upsilon=1}^rp_i=1 \text{ and } \log p \in \text{rowspan}(A)\}$ which is strongly related to $I_A$ as Theorem~\ref{thm:Fundamental_thm} below demonstrates. Consequently, we say that $A$ serves as the \emph{design matrix} of the model $\mathcal M_A$ or equivalently the design matrix of the toric ideal $I_A$. 

\begin{thm}\label{thm:Fundamental_thm} (Fundamental Theorem of Markov Bases \cite[Theorem 3.1]{DS98}) Let $A$ be a integer matrix. A finite set of moves $\mathcal B\subset \ker_\Z A$ is a Markov basis for $A$ if the corresponding set of binomials $\{x^{m^+}-x^{m^-}:m\in \mathcal B\}$ is a generating set for the toric ideal $I_A$.
\end{thm}

Furthermore,  \cite[Theorem 3.2]{DS98}  provides an algorithm to compute a generating set for $I_A$ using the elimination theorem and Gr\"obner bases,  thereby offering a theoretical solution for the computation of Markov bases. However, even the state-of-the-art  algorithms for computing Markov bases, such as  \cite{HemmeckeMalkin2009}, face practical   limitations: the computations do not scale well.  Hence, a combinatorial description of Markov bases with variable parameters---such as the number of vertices and colors in our case, as presented in Theorem~\ref{thm:generators}---is highly desirable. Further details and references on Markov bases can be found in \cite{SethBook},  \cite{AHT2012}, and \cite{petrovic2019}, while  \cite{MarkovBases25years} provides a recent overview of the state of the art.

\smallskip 
To provide context for the final contribution of this paper, we consider the perspective of discrete geometry. 
Namely,  the leading terms of a Gr\"obner basis of $I_A$ correspond to the minimal non-faces determining a triangulation for the columns of the matrix $A$  \citep[Chapter 9.4]{DeLoeraetal2010triangulations}. 
The fact that a Gr\"obner bases requires the specification of a monomial order further results  
in a unique-sink orientation of the graph with vertices in the fiber $\mathcal F_A(b)$. This property has important computational consequences, as shown for example in the  early works of \cite{Sturmfels91Toric}, \cite{Thomas95GeometricBuchberger}, and \cite{DeLoera1995hypersimplex}. In the latter, Gr\"obner bases were used to study triangulations and optimization of the $b$-matching problem--the graph with the smallest cost having a given degree sequence $b$. The final contribution of this paper generalizes the quadratic Gr\"obner basis result from the monochromatic case proved in \cite{DeLoera1995hypersimplex} to the following general case. 

\begin{restatable}{thm}{thmthree}\label{thm:Grobner_basis} There exists a monomial order $\succ$ on $K[x_{uv}:1\leq u<v\leq n]$ such that for any $k$-coloring $z$ of $[n]$, the set $\{x^{\gamma^+}-x^{\gamma^-}: \gamma\in {\mathcal{M}_{n,z}}\}$ is in fact a Gr\"obner basis for $\IDC$ with respect to $\succ$.
\end{restatable}

The monomial order referred to in Theorem \ref{thm:Grobner_basis} is introduced at the beginning of Section~\ref{sec:grobner}. Its origin lies in geometric considerations, which will be essential for the proof of this result.

\section{A quadratic Markov basis for the $\beta$-SBM matrix}

Let us begin by noting that any vector $\gamma\in \Z^{\binom{n}{2}}$ can be interpreted as a graph by defining $V(\gamma)=[n]$ as the set of vertices and $E(\gamma)=\{\{u,v\}:\gamma_{uv}\neq 0\}$ as the set of edges. Furthermore, we say that an edge $uv\in E(\gamma)$ is \emph{positive} if $\gamma_{uv}>0$ and \emph{negative} if $\gamma_{uv}<0$, with $|\gamma_{uv}|$ representing the multiplicity of the edge $uv$. Here $\gamma_{uv}$ denotes the entry in $\gamma$ asociated to the 2-subset $\{u,v\}$ of $[n]$, and as such, we use $\gamma_{uv}$ or $\gamma_{vu}$ interchangeably. For simplicity we may occassionally abuse  notation and treat $\gamma\in \Z^{\binom{n}{2}}$ as a graph, given the appropriate context. 

For $\gamma\in \Z^{\binom{n}{2}}$ and $v\in [n]$ we define the \emph{positive degree} and \emph{negative degree} of $v$ as 
\begin{equation}\label{eqn:pos-neg-degree}
\deg_\gamma^+(v):=\sum_{u\in [n]: \gamma_{uv}>0}\gamma_{uv}\;\; \text{ and } \;\;\deg_\gamma^-(v):=\sum_{u\in [n]: \gamma_{uv}<0}-\gamma_{uv},
\end{equation}
respectively. In other words, $deg^+_\gamma(u)$ and $deg^-_\gamma(u)$ are the numbers of positive and negative edges incident with $u$, respectively. We define the \emph{positive degree sequence} and the \emph{negative degree sequence} as the vectors $d^+(\gamma)=(\deg_\gamma^+(1),\ldots ,\deg_\gamma^+(n))$  and $d^-(\gamma)=(\deg_\gamma^-(1),\ldots ,\deg_\gamma^-(n))$, respectively. Also, for any $1\leq i\leq j\leq k$, let
\begin{equation}\label{eqn:pos-neg-color}
c^+_\gamma(z,i,j):=\sum_{\substack{u\in z^{-1}(i),\; v\in z^{-1}(j): \\ \gamma_{uv}>0}} \gamma_{uv} \;\;\text{ and }\;\; c^-_\gamma(z,i,j):=\sum_{\substack{u\in z^{-1}(i),\; v\in z^{-1}(j): \\ \gamma_{uv}<0}} -\gamma_{uv}.
\end{equation}
This means that $c^+_\gamma(z,i,j)$ and $c^-_\gamma(z,i,j)$ are the number of positive and negative edges, respectively, that are connecting an $i$-th colored vertex with an $j$-th colored vertex. We define the \emph{positive color sequence} and the \emph{negative color sequence} as the vectors $c^+(z,\gamma)=\big(c^+_{\gamma}(z,i,j): 1\leq i\leq j\leq k\big)$ and $c^-(z,\gamma)=\big(c^-_{\gamma}(z,i,j): 1\leq i\leq j\leq k\big)$, respectively. Notice that when $\gamma\in \N^{\binom{n}{2}}$, $(d^+(\gamma); c^+(z,\gamma))$ coincides with the $c$-degree sequence defined in the introduction and $(d^-(\gamma); c^-(z,\gamma))$ is a vector of zeros. When $z$ is clear from the context we will write $c_\gamma^{\pm}(i,j)$ instead of $c^{\pm}_\gamma(z,i,j)$. We say that $\gamma\in \Z^{\binom{n}{2}}$ satisfies the \emph{degree-balance condition} if $d^+(\gamma)=d^-(\gamma)$ and the \emph{color-balance condition with respect to $z$} if $c^+(z,\gamma)=c^-(z,\gamma)$. When $z$ is clear from the context, we simply say that $\gamma$ satisfies the color-balance condition.

Let $n\in \Z_+$ be a positive integer, $z$ a $k$-coloring of $[n]$, and $\DC_{\!n,z}$ the design matrix of the $\beta$-SBM. Notice that for any $\gamma\in \Z^{\binom{n}{2}}$, 
\begin{equation}\label{eqn:matrix_mult}
\DC_{\!n,z}\!\gamma = (d^+(\gamma)-d^-(\gamma); c^+(\gamma)-c^-(\gamma)).
\end{equation}

This means that $\gamma\in \ker_{\Z}\DC_{\!n,z}$ if and only if $\gamma$ satisfies both the degree and the color-balance conditions. It is not hard to see that any $\gamma\in \Z^{\binom{n}{2}}$ satisfying the degree-balance condition must be a union of closed even walks, such that the edges in the walk alternate between positive and negative edges. Here, a walk of length $k$ on a graph is a sequence of vertices and edges $\{v_1,e_1,v_2,e_2,\dots,v_k,e_k,v_{k+1}\}$; it is said to be closed if $v_{k+1}=v_1$, and even if $k$ is an even integer. 
Furthermore, when $\DC_{\!n,z}$ is regarded as the incidence graph of a 3-uniform hypergraph, the elements of $\ker_\Z \!\DC_{\!n,z}$ can be understood using the notion that generalizes closed even walks on a graph to a type of walk on a hypergraph that preserves the balancing condition. Namely, such constructions on hypergraphs are known as \emph{monomial walks} in the literature (\cite{PShypergraphs}, \cite{Vill}). Hence we say that $\gamma\in \Z^{\binom{n}{2}}$ is a \emph{monomial walk with respect to $z:[n]\to[k]$} if $\gamma\in \ker_\Z\!\DC_{\!n,z}$. For convenience, when considering a monomial walk $\gamma$, sometimes we will describe it using a vertex sequence enclosed by square brackets: $[v_1,v_2, \ldots, v_{2l-1}, v_{2l}]$. This notation means that $\gamma$ is the element in $\ker_\Z\!\DC_{\!n,z}$ entrywise defined by
\[
\gamma_{uv}=\sum_{i=1}^l\mathds{1}_{\{u,v\}=\{v_{2i-1},v_{2i}\}}-\mathds{1}_{\{u,v\}=\{v_{2i},v_{2i+1}\}},
\]
where $2l+1=1$ and for an statement $P$,
\[
   \mathds{1}_P=
   \begin{cases}
      1 & \text{ if } P \text{ is true,}\\
      0 & \text{otherwise}.
   \end{cases}
\]
The use of the different notations (either vector or brackets) will depend on the context.

\begin{example}\label{example:monomial_walk}
Let $\DC_{5,z}$ the matrix of Example \ref{example:matrix} and $\gamma=(0,0,-1,1,-1,2,-1,0,1,-1)\in \ker_\Z\!\DC_{n,5}$ be the monomial walk illustrated below. One way to write $\gamma$ using bracket notation is $\gamma=[1,5,2,4,5,3,2,4]$. In this case $x^{\gamma^+}-x^{\gamma^-}=x_{15}x_{24}^2x_{35}-x_{14}x_{23}x_{25}x_{45}$. 

\begin{figure}[h]
\centering
\begin{tikzpicture}[scale=1, every node/.style={minimum size=5mm, text centered}]
   
   % Draw the circular graph
   \foreach \i/\col in {1/Bcolor, 2/Bcolor, 3/Rcolor, 4/Rcolor, 5/Gcolor} {
       \node[draw, circle, preaction={fill=\col, fill opacity=0.1}, draw=\col, inner sep=0pt] (n\i) at ({(-\i)*360/5+72}:1.75) {\footnotesize{\i}};
   }
   
   % Draw the edges
    \path
      (n2) edge[bend left=17] (n4)
      (n4) edge[bend left=17] (n2)
      (n1) edge (n5)
      (n3) edge (n5)
      (n2) edge[dashed] (n3)
      (n2) edge[dashed] (n5)
      (n4) edge[dashed] (n5)
      (n1) edge[dashed] (n4);
\end{tikzpicture}
\caption{Monomial walk $m=[1,5,2,4,5,3,2,4]$.} 
\label{fig:mon_walk}
\end{figure}
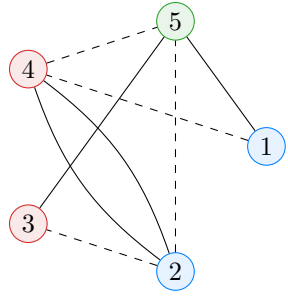

\end{example}

As noticed previously, for any color-graphical $c$-degree sequence $(d;c)\in \N^{n+\binom{k+1}{2}}$, $\fiberDC(d;c)$ is the set of multigraphs with fixed $c$-degree sequence $(d;c)$. Meaning that a Markov basis for $\DC_{\!n,z}$ is a set of moves $\mathcal B$ that allow us to connect any two multigraphs with a fixed $c$-degree sequence by using moves in $\mathcal B$. To prove Theorem~\ref{thm:generators} we use the algebraic analogue provided by Theorem~\ref{thm:Fundamental_thm}. In other words, we show that $\mathcal M_{n,z}=\{\gamma\in \ker_\Z\!\DC_{\!n,z}: ||\gamma||_1=4\}$ is a Markov basis for $\DC_{\!n,z}$ by proving that $I_{\mathcal M_{n,z}}:=\langle\{x^{\gamma^+}-x^{\gamma^-}:\gamma\in \mathcal M_{n,z}\}\rangle$ is equal to $\IDC$.

To do so, we will use the combinatorial conditions of monomial walks in order to reduce binomials of degree greater than two by splitting the monomial walks of length longer than four into shorter walks. First, let us see that the toric monomial map \eqref{eqn:toricmap} whose vanishing ideal is $\IDC$ can be explicitly written as
\begin{align}\label{eqn:toricmap_betaSBM}
\notag 
    \varphi_\beta: K[x_{uv} : 1\leq u<v\leq n] &\rightarrow K[\{s_1,\ldots, s_n\}\cup \{t_{ij}:1\leq i\leq j\leq k\}];\\
     x_{uv} & \mapsto s_us_v t_{z(u)z(v)}. 
\end{align}

Then, for any $\gamma\in \Z^{\binom{n}{2}}$, $\vphi_\beta(x^{\gamma^+})=\vphi_\beta(x^{\gamma^-})$ if and only if $\DC_{\!n,z}\!\gamma^+=\DC_{\!n,z}\!\gamma^-$, which means that $x^{\gamma^+}-x^{\gamma^-}\in \ker \vphi_\beta= \IDC$ if and only if $\gamma\in \ker_\Z\!\DC_{\!n,z}$. In fact, as a consequence of \cite[Corollary 4.3]{St} it follows that $\IDC=\langle \{x^{\gamma^+}-x^{\gamma^-}: \gamma\in \ker_{\Z} \DC_{\!n,z}\}\rangle$. This immediately implies that $I_{\mathcal M_{n,z}}\subseteq \IDC$. Therefore, to prove Theorem~\ref{thm:generators} it suffices to show that $\IDC\subseteq I_{\mathcal M_{n,z}}$. To do so, we start by providing a combinatorial description of $\mathcal M_{n,z}$.

\begin{lemma}\label{lemma:4cycles} 
$\mathcal M_{n,z}=\{[uvu'v']: z(u)=z(u') \text{ or } z(v)=z(v')\}$. In other words, the elements of $\mathcal M_{n,z}$ are 4-cycles with at least two opposite vertices of the same color. 
\end{lemma}

\begin{proof} First suppose without losing generality that $\gamma=[uvu'v']$ with $z(u)=z(u')$. By convention this means that $uv, u'v'$ are positive edges, $vu', v'u$ are negative edges in $E(\gamma)$. Since $z(u)=z(u')$, we have that $t_{z(u)z(v)}=t_{z(u')z(v)}$ and $t_{z(u')z(v')}=t_{z(u)z(v')}$, which means
\begin{align*}
\vphi_\beta(x^{\gamma^+}-x^{\gamma^-})&=s_us_vt_{z(u)z(v)}\cdot s_{u'}s_{v'}t_{z(u')z(v')}-s_us_{v'}t_{z(u)z(v')}\cdot s_{u'}s_{v}t_{z(u')z(v)}\\
&= s_us_vs_{u'}s_{v'}(t_{z(u)z(v)}t_{z(u')z(v')}-t_{z(u)z(v')}t_{z(u')z(v)})=0, 
\end{align*}
implying that $x^{\gamma^+}-x^{\gamma^-}\in \IDC$, or equivalenty, $\gamma\in \mathcal M_{n,z}$.

Now, suppose that $\gamma\in \mathcal M_{n,z}$. The degree-balance condition implies that $\gamma$ is a 4-cycle $[uvu'v']$ with $uv, u'v'$ positive and $vu', v'u$ negative edges. Because $\gamma$ has a positive edge between colors $z(u)$ and $z(v)$, the color-balance condition implies that at least one of the negative edges $uv'$ or $u'v$ connects colors $z(u)$ and $z(v)$. This is equivalent to having $z(v')=z(v)$ or $z(u')=z(u)$.
\end{proof}

Showing that $\IDC\subseteq I_{\mathcal M_{n,z}}$ relies on the fact that for any $x^{\gamma^+}-x^{\gamma^-}\in \IDC$, we can either peel off $4$-cycles from $\gamma$ (Lemma \ref{lemma:reduce}) or, alternatively, use $4$-cycles to reconnect $\gamma$ (Lemma \ref{lemma:good2walk}). This allows us to obtain a new monomial walk from which we can peel off $4$-cycles belonging to $\IDC$. This process enables us to express $x^{\gamma^+}-x^{\gamma^-}$ as the sum of an element in $I_{\mathcal M_{n,z}}$ and a binomial in $\IDC$ with a degree smaller than $\deg(x^{\gamma^+}-x^{\gamma^-})$. Before presenting the proofs, we illustrate the idea with the following example.

\begin{example} Let $n=8, k=2$ and let $\{\{1,4,5,6\}, \{2,3,7,8\}\}$ be the partition induced by $z$, which assigns colors \begin{tikzpicture}[every node/.style={minimum size=.3mm}] \node[draw, circle, draw=Bcolor, fill=Bcolor, fill opacity=0.2, inner sep=1.8pt] (x) at (0,0){};\end{tikzpicture} and \begin{tikzpicture}[every node/.style={minimum size=.3mm}] \node[draw, circle, draw=Rcolor, fill=Rcolor, fill opacity=0.2, inner sep=1.8pt] (x) at (0,0){};\end{tikzpicture}, respectively. Consider $f=x^{\gamma^+}-x^{\gamma^-}=x_{12}x_{34}x_{56}x_{78}-x_{14}x_{23}x_{58}x_{67}\in \IDC$ and observe that $f=x_{34}x_{56}(x_{12}x_{78}-x_{17}x_{28})+f'$ where $f'=x^{\gamma'^+}-x^{\gamma'^-}=x_{34}x_{56}x_{17}x_{28}-x_{14}x_{23}x_{58}x_{67}$. As we illustrate in the picture below, this way of rewriting $f$ corresponds to rewriting $\gamma$ as a sum of a 4-cycle and a monomial walk $\gamma'$ with the same number of edges as $\gamma$. Notice that $\gamma$ and $\gamma'$ differ by a switch, i.e., an exchange of two edges by another two edges on the same quadruple of vertices.

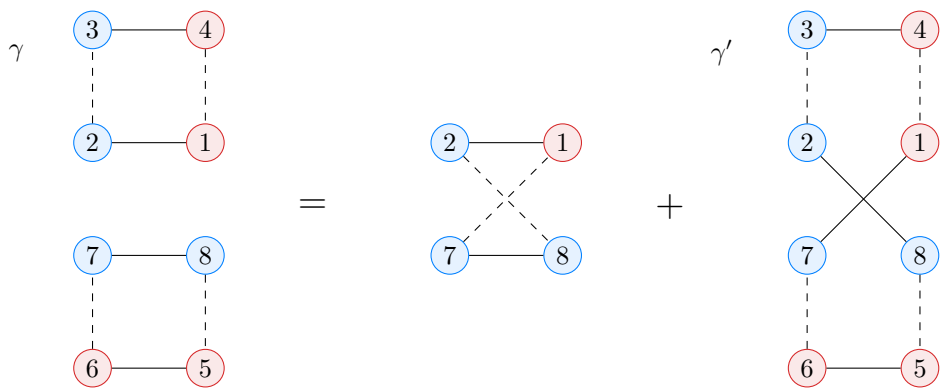
\begin{figure}[h]
\centering
\begin{tikzpicture}[scale=1, every node/.style={minimum size=5mm, text centered}]
  
  % \tikzset{Bullet/.style={fill=black,draw,color=#1,circle,minimum size=1pt,scale=0.55}}

  % Original walk
  \node[draw, circle, preaction={fill=Rcolor, fill opacity=0.1}, draw=Rcolor, inner sep=0pt] (1) at (0,0) {$1$};

  \node[draw, circle, preaction={fill=Bcolor, fill opacity=0.1}, draw=Bcolor, inner sep=0pt] (2) at (-1.5,0) {$2$};
  \node[draw, circle, preaction={fill=Bcolor, fill opacity=0.1}, draw=Bcolor, inner sep=0pt] (3) at (-1.5,1.5) {$3$};
  \node[draw, circle, preaction={fill=Rcolor, fill opacity=0.1}, draw=Rcolor, inner sep=0pt] (4) at (0,1.5) {$4$};
  
  \node[draw, circle, preaction={fill=Bcolor, fill opacity=0.1}, draw=Bcolor, inner sep=0pt] (8) at (0,-1.5) {$8$};
  \node[draw, circle, preaction={fill=Bcolor, fill opacity=0.1}, draw=Bcolor, inner sep=0pt] (7) at (-1.5,-1.5) {$7$};
  \node[draw, circle, preaction={fill=Rcolor, fill opacity=0.1}, draw=Rcolor, inner sep=0pt] (5) at (0,-3) {$5$};
  \node[draw, circle, preaction={fill=Rcolor, fill opacity=0.1}, draw=Rcolor, inner sep=0pt] (6) at (-1.5,-3) {$6$};

  %Equal sign
  \node[label={left, yshift=-.1cm: \large =}] (eq_sign) at (2,-.75) {};  

  %4-cycle
  \node[draw, circle, preaction={fill=Rcolor, fill opacity=0.1}, draw=Rcolor, inner sep=0pt] (1_right) at (4.75,0) {$1$};
  \node[draw, circle, preaction={fill=Bcolor, fill opacity=0.1}, draw=Bcolor, inner sep=0pt] (2_right) at (3.25,0) {$2$};
  \node[draw, circle, preaction={fill=Bcolor, fill opacity=0.1}, draw=Bcolor, inner sep=0pt] (8_right) at (4.75,-1.5) {$8$};
  \node[draw, circle, preaction={fill=Bcolor, fill opacity=0.1}, draw=Bcolor, inner sep=0pt] (7_right) at (3.25,-1.5) {$7$};

  %Plus sign
  \node[label={left, yshift=-.1cm: \large +}] (plus_sign) at (6.75,-.75) {};

  %Reconnection
  \node[draw, circle, preaction={fill=Rcolor, fill opacity=0.1}, draw=Rcolor, inner sep=0pt] (1_rightt) at (9.5,0) {$1$};
  \node[draw, circle, preaction={fill=Bcolor, fill opacity=0.1}, draw=Bcolor, inner sep=0pt] (2_rightt) at (8,0) {$2$};
  \node[draw, circle, preaction={fill=Bcolor, fill opacity=0.1}, draw=Bcolor, inner sep=0pt] (3_rightt) at (8,1.5) {$3$};
  \node[draw, circle, preaction={fill=Rcolor, fill opacity=0.1}, draw=Rcolor, inner sep=0pt] (4_rightt) at (9.5,1.5) {$4$};
  
  \node[draw, circle, preaction={fill=Bcolor, fill opacity=0.1}, draw=Bcolor, inner sep=0pt] (8_rightt) at (9.5,-1.5) {$8$};
  \node[draw, circle, preaction={fill=Bcolor, fill opacity=0.1}, draw=Bcolor, inner sep=0pt] (7_rightt) at (8,-1.5) {$7$};
  \node[draw, circle, preaction={fill=Rcolor, fill opacity=0.1}, draw=Rcolor, inner sep=0pt] (5_rightt) at (9.5,-3) {$5$};
  \node[draw, circle, preaction={fill=Rcolor, fill opacity=0.1}, draw=Rcolor, inner sep=0pt] (6_rightt) at (8,-3) {$6$};

  %Labels
  \node[label={left, yshift=-.1cm: \textcolor{black}{$\gamma$}}] (gamma_f) at (-2,1.3) {};
  % \node[label={left, yshift=-.1cm: \textcolor{gray}{$[1,2,7,8]$}}] (cycle) at (4.75,1) {};
  \node[label={left, yshift=-.1cm: \textcolor{black}{$\gamma'$}}] (gamma_f') at (7.4,1.3) {};

  \path
  (1) edge (2)
  (2) edge[dashed] (3)
  (3) edge (4)
  (4) edge[dashed] (1)
  (7) edge (8)
  (8) edge[dashed] (5)
  (5) edge (6)
  (6) edge[dashed] (7)
  (1_right) edge (2_right)
  (2_right) edge[dashed] (8_right)
  (8_right) edge (7_right)
  (7_right) edge[dashed] (1_right)
  (1_rightt) edge (7_rightt)
  (2_rightt) edge[dashed] (3_rightt)
  (3_rightt) edge (4_rightt)
  (4_rightt) edge[dashed] (1_rightt)
  (2_rightt) edge (8_rightt)
  (8_rightt) edge[dashed] (5_rightt)
  (5_rightt) edge (6_rightt)
  (6_rightt) edge[dashed] (7_rightt);
\end{tikzpicture}
\caption{Reconnecting with a 4-cycle. On the left of the equality is the graph $\gamma$, while on the right are the $4$-cycle corresponding to $x_{12}x_{78}-x_{17}x_{28}$ and $\gamma'$.} 
\label{fig:reconnect}
\end{figure}

Now, we can peel off a $4$-cycle from $\gamma'$ to obtain a monomial walk $\gamma''$ with less edges than $\gamma'$ (see Figure~\ref{fig:peel_off}). Algebraically, this means that $f'=x_{23}x_{58}(x_{17}x_{46}-x_{14}x_{67})+x_{17}(x_{28}x_{34}x_{56}-x_{23}x_{46}x_{58})$, where $f''=x^{\gamma''^+}-x^{\gamma''^-}=x_{28}x_{34}x_{56}-x_{23}x_{46}x_{58}$. In fact, we can continue peeling off $4$-cycles from $\gamma''$ in order to prove that $f\in \langle x^{\gamma^+}-x^{\gamma^-}:\gamma\in \mathcal M_{n,z}\rangle$. 

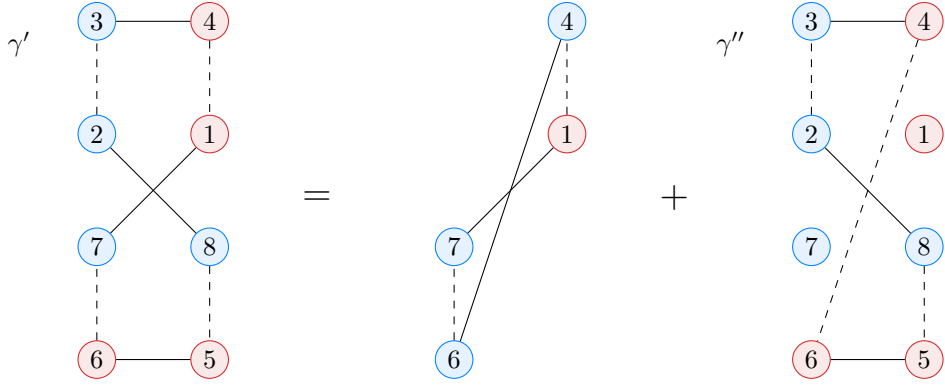
\begin{figure}[h]
\centering
\begin{tikzpicture}[scale=1, every node/.style={minimum size=5mm, text centered}]
  
% Original walk
  \node[draw, circle, preaction={fill=Rcolor, fill opacity=0.1}, draw=Rcolor, inner sep=0pt] (1) at (0,0) {$1$};

  \node[draw, circle, preaction={fill=Bcolor, fill opacity=0.1}, draw=Bcolor, inner sep=0pt] (2) at (-1.5,0) {$2$};
  \node[draw, circle, preaction={fill=Bcolor, fill opacity=0.1}, draw=Bcolor, inner sep=0pt] (3) at (-1.5,1.5) {$3$};
  \node[draw, circle, preaction={fill=Rcolor, fill opacity=0.1}, draw=Rcolor, inner sep=0pt] (4) at (0,1.5) {$4$};
  
  \node[draw, circle, preaction={fill=Bcolor, fill opacity=0.1}, draw=Bcolor, inner sep=0pt] (8) at (0,-1.5) {$8$};
  \node[draw, circle, preaction={fill=Bcolor, fill opacity=0.1}, draw=Bcolor, inner sep=0pt] (7) at (-1.5,-1.5) {$7$};
  \node[draw, circle, preaction={fill=Rcolor, fill opacity=0.1}, draw=Rcolor, inner sep=0pt] (5) at (0,-3) {$5$};
  \node[draw, circle, preaction={fill=Rcolor, fill opacity=0.1}, draw=Rcolor, inner sep=0pt] (6) at (-1.5,-3) {$6$};

  %Equal sign
  \node[label={left, yshift=-.1cm: \large =}] (eq_sign) at (2,-.75) {};  

  %4-cycle
  \node[draw, circle, preaction={fill=Rcolor, fill opacity=0.1}, draw=Rcolor, inner sep=0pt] (1_right) at (4.75,0) {$1$};
  \node[draw, circle, preaction={fill=Bcolor, fill opacity=0.1}, draw=Bcolor, inner sep=0pt] (4_right) at (4.75,1.5) {$4$};
  \node[draw, circle, preaction={fill=Bcolor, fill opacity=0.1}, draw=Bcolor, inner sep=0pt] (6_right) at (3.25,-3) {$6$};
  \node[draw, circle, preaction={fill=Bcolor, fill opacity=0.1}, draw=Bcolor, inner sep=0pt] (7_right) at (3.25,-1.5) {$7$};

  %Plus sign
  \node[label={left, yshift=-.1cm: \large +}] (plus_sign) at (6.75,-.75) {};

  %Reconnection
  \node[draw, circle, preaction={fill=Rcolor, fill opacity=0.1}, draw=Rcolor, inner sep=0pt] (1_rightt) at (9.5,0) {$1$};
  \node[draw, circle, preaction={fill=Bcolor, fill opacity=0.1}, draw=Bcolor, inner sep=0pt] (2_rightt) at (8,0) {$2$};
  \node[draw, circle, preaction={fill=Bcolor, fill opacity=0.1}, draw=Bcolor, inner sep=0pt] (3_rightt) at (8,1.5) {$3$};
  \node[draw, circle, preaction={fill=Rcolor, fill opacity=0.1}, draw=Rcolor, inner sep=0pt] (4_rightt) at (9.5,1.5) {$4$};
  
  \node[draw, circle, preaction={fill=Bcolor, fill opacity=0.1}, draw=Bcolor, inner sep=0pt] (8_rightt) at (9.5,-1.5) {$8$};
  \node[draw, circle, preaction={fill=Bcolor, fill opacity=0.1}, draw=Bcolor, inner sep=0pt] (7_rightt) at (8,-1.5) {$7$};
  \node[draw, circle, preaction={fill=Rcolor, fill opacity=0.1}, draw=Rcolor, inner sep=0pt] (5_rightt) at (9.5,-3) {$5$};
  \node[draw, circle, preaction={fill=Rcolor, fill opacity=0.1}, draw=Rcolor, inner sep=0pt] (6_rightt) at (8,-3) {$6$};

  %Labels
  \node[label={left, yshift=-.1cm: \textcolor{black}{$\gamma'$}}] (gamma_f) at (-2,1.3) {};
  % \node[label={left, yshift=-.1cm: \textcolor{gray}{$\Gamma_{x_{17}x_{46}-x_{14}x_{67}}$}}] (cycle) at (4.5,1.2) {};
  \node[label={left, yshift=-.1cm: \textcolor{black}{$\gamma''$}}] (gamma_f') at (7.5,1.3) {};

  \path
  (1) edge (7)
  (2) edge[dashed] (3)
  (3) edge (4)
  (4) edge[dashed] (1)
  (2) edge (8)
  (8) edge[dashed] (5)
  (5) edge (6)
  (6) edge[dashed] (7)
  (1_right) edge (7_right)
  (1_right) edge[dashed] (4_right)
  (4_right) edge (6_right)
  (6_right) edge[dashed] (7_right)
  (2_rightt) edge[dashed] (3_rightt)
  (3_rightt) edge (4_rightt)
  (4_rightt) edge[dashed] (6_rightt)
  (2_rightt) edge (8_rightt)
  (8_rightt) edge[dashed] (5_rightt)
  (5_rightt) edge (6_rightt);
\end{tikzpicture}
\caption{Peeling off a 4-cycle. On the left of the equality is the graph $\gamma'$, while on the right are the $4$-cycle corresponding to $x_{17}x_{46}-x_{14}x_{67}$ and $\gamma''$.} 
\label{fig:peel_off}
\end{figure}
    
\end{example}

\begin{lemma}\label{lemma:good2walk} 
For any $f=x^{\gamma^+}-x^{\gamma^-}\in \IDC$, there exists $f'\in I_{{\mathcal{M}_{n,z}}}$ and $f''\in \IDC$ with $\deg(f)=\deg(f'')$ such that $f=f'+f''$, where $f''=x^{\gamma''^+}-x^{\gamma''^-}$ and $\gamma''$ contains a subwalk $uvw$ such that $z(u)=z(w)$ and $uv, vw$ have different signs.
\end{lemma}

\begin{proof} Let $f=x^{\gamma^+}-x^{\gamma^-}\in \IDC$ and let $uv$ be a positive edge in $\gamma$. By the color-balance condition, there must exists a negative edge $u'v'\in E(\gamma)$ such that $z(u')=z(u)$ and $z(v')=z(v)$. Since no edge in $\gamma$ can be positive and negative at the same time, it follows that $uv\neq u'v'$. Moreover, if $u=u'$ then $vuv'$ is a subwalk such that $z(v)=z(v')$ with $vu$ positive and $uv'$ negative so the statement would follow. Hence, we assume that $u\neq u'$ and similarly $v\neq v'$.   

Now, given that each of $u'$ and $v'$ are adjacent to the negative edge $u'v'$, the degree-balance condition guarantees the existence of positive edges $u'\hat{w}, v'w$ in $E(\gamma)$. Let us consider the following $2$ cases:

\begin{enumerate}[label=(\roman*)]
    \item $\{w,\hat{w}\}=\{u,v\}$. 

    In case $w=u, \hat{w}=v$, then $uv'u'$ is a subwalk such that $z(u)=z(u')$, $uv'=wv'$ is positive and $v'u'$ is negative, so it would be enough to take $f'=0, f''=f$. Then, assume that $w=v, \hat{w}=u$. In such case, since $u$ is adjacent to two positive edges, there must exists (by the degree-balance condition) a negative edge $u\hat{u}\in E(\gamma)$. Notice that since $z(u)=z(u')$, $x_{u'\hat{u}}x_{uv'}-x_{u\hat{u}}x_{u'v'}\in \IDC$ and 
    \begin{align*}
    f=x^{\gamma^+}-x^{\gamma^-}&=x^{\gamma^+}-x^{\alpha}x_{u\hat{u}}x_{u'v'}\\
    &=x^\alpha(x_{u'\hat{u}}x_{uv'}-x_{u\hat{u}}x_{u'v'})+(x^{\gamma^+}-x^\alpha x_{u'\hat{u}}x_{uv'}),
    \end{align*}
    where, $\alpha\in \N^{\binom{n}{2}}$ is such that $x^\alpha x_{u\hat{u}}x_{u'v'}=x^{\gamma^-}$. Then we take $f'=x^\alpha(x_{u'\hat{u}}x_{uv'}-x_{u\hat{u}}x_{u'v'})$, $f''=x^{\gamma^+}-x^\alpha x_{u'\hat{u}}x_{uv'}$ and the subwalk $vuv'$ satisfies that $z(v)=z(v')$, $uv$ is positive and $uv'$ is negative.

    \item $\{w,\hat{w}\}\neq\{u,v\}$.

    Assume without loss of generality that $w\notin \{u,v\}$. Given that $z(v)=z(v')$, it follows that $x_{uv}x_{wv'}-x_{uv'}x_{wv}\in \IDC$. Let us observe that
    \begin{align*}
    f=x^{\gamma^+}-x^{\gamma^-}&=x^\alpha x_{uv}x_{wv'}-x^{\gamma^-}\\
    &=x^\alpha(x_{uv}x_{wv'}-x_{uv'}x_{wv})+(x^\alpha x_{uv'}x_{wv}-x^{\gamma^-}), 
    \end{align*}
    where $\alpha\in\N^{\binom{n}{2}}$ is such that $x^\alpha x_{uv}x_{wv'}=x^{\gamma^+}$. In this case we set $f'=x^\alpha(x_{uv}x_{wv'}-x_{uv'}x_{wv})$, $f''=x^\alpha x_{uv'}x_{wv}-x^{\gamma^-}\in \IDC$. Notice that $\deg(f'')=\deg(f)$ and that the subwalk $uv'u'$ satisfies that $z(u)=z(u')$, $uv'$ is positive and $v'u'$ is negative.    
    
\end{enumerate}

\end{proof}

\begin{lemma}\label{lemma:reduce}
Let $f=x^{\gamma^+}-x^{\gamma^-}\in \IDC$. Suppose $\gamma$ contains a subwalk $uvw$ such that $z(u)=z(w)$ and $uv, vw$ have different signs. Then, $f=f'+x_{uv}f''$ for some $f'\in I_{\mathcal M_{n,z}}$ and $f''\in \IDC$ with $\deg(f'')\leq \deg(f)-1$.    
\end{lemma}

\begin{proof}
Suppose without loss of generality that $uv$ is negative and $vw$ is positive. Then the degree-balance condition guarantees the existence of a positive edge $uu'$. Consider the following two cases:
\begin{enumerate}[label=(\roman*)]
    \item $u'\neq w$.

    Since $z(u)=z(w)$, it follows that $x_{uu'}x_{wv}-x_{uv}x_{wu'}\in \IDC$. Let $\alpha, \alpha'\in \N^{\binom{n}{2}}$ be such that $x^{\gamma^+}=x^\alpha x_{uu'}x_{wv}$ and $x^{\gamma^-}=x_{uv}x^{\alpha'}$. Then, we have that
    \begin{align*}
    f = x^{\gamma^+}-x^{\gamma^-}&=x^\alpha x_{uu'}x_{wv}-x_{uv}x^{\alpha'}\\
    &=x^\alpha(x_{uu'}x_{wv}-x_{uv}x_{wu'})+x_{uv}(x^{\alpha}x_{wu'}-x^{\alpha'}).
    \end{align*}
    Let $f'=x^\alpha(x_{uu'}x_{wv}-x_{uv}x_{wu'})$ and $f''=x^\alpha x_{wu'}-x^{\alpha'}$. Given that $\IDC$ is prime and $f,x_{uu'}x_{wv}-x_{uv}x_{wu'}$ both belong to $\IDC$ we have that $f''\in \IDC$. Furthermore, either $f''=0$ or $\deg(f'')=\deg(f)-1$.

    \item $u'=w$.

    In this case $uw$ is a positive edge in $\gamma$ so by the degree-balance condition there must be a negative edge $ww'$ with $w'\notin\{u,v\}$. This situation is analogous to the previous case since $uvw$ is a subwalk with $z(u)=z(w)$, $uv$ negative, $vw$ positive and $ww'$ a negative edge in $\gamma$ such that $w'\neq u$.  
\end{enumerate}
\end{proof}

\thmone*

% \begin{proof}[Proof of Theorem \ref{thm:generators}]
\begin{proof}
As previously mentioned, it suffices  to show that $\IDC\subseteq I_{\mathcal M_{n,z}}$. Let us remember that $\IDC=\langle x^{\gamma^+}-x^{\gamma^-}:\gamma\in \ker_\Z\!\DC_{\!n,z}\rangle$ and let $f=x^{\gamma^+}-x^{\gamma^-}\in \IDC$. If $\deg(f)=2$, then $f\in I_{\mathcal{M}_{n,z}}$ by definition. Suppose that $\deg(f)=k+1$. By Lemma~\ref{lemma:good2walk} and Lemma~\ref{lemma:reduce} we can write $f=f'+x_{uv}f''$ for some $1\leq u<v\leq n$ where $f'\in I_{\mathcal{M}_{n,z}}$ and $\deg(f'')\leq k$. By induction on the degree we have that $f''\in I_{\mathcal{M}_{n,z}}$, hence $f\in I_{\mathcal{M}_{n,z}}$.
\end{proof}

Theorem~\ref{thm:generators} has a direct application in the construction of a goodness-of-fit for exponential families having the $c$-degree vector as their sufficient statistics as we explain below.

Let $\mathbb G\subset \N^{\binom{n}{2}}$ be the space of graphs on $n$ colored vertices, where we think of vertices as being are partitioned into blocks by a fixed vertex coloring $z:[n]\to[k]$.  It turns out that this $\mathbb G$ is a sample space for the following exponential family of distributions. Following  \cite{karwa2016exact}, we say that a family of probability distributions $\{p_{\alpha, \beta}: (\alpha, \beta)=((\alpha_{ij})_{1\leq i \leq j \leq k}, (\beta_u)_{u=1}^n )\in \Theta\}$ over $\mathbb{G}$ with parameter space $\Theta\subset\mathbb{R}^{\binom{k+1}{2}}\times \mathbb{R}^n$ is a \emph{generalized $\beta$-SBM} if the distributions are of the following form for all  $\gamma \in \mathbb{G}$: 
\[
p_{\alpha, \beta}(\gamma)=\frac{h(\gamma)}{\psi(\alpha, \beta)}\exp\Big\{(\beta,\alpha)^\top\!\!\DC_{\!n,z}\!\gamma\Big\} ,%\;\;\forall \gamma \in \mathbb{G},
\] 
for some  $(\alpha,\beta)\in\Theta$. 
Here, $h$ represents the {base measure} and $\psi(\alpha, \beta)$ is the normalizing constant of the family. %Idk if I should add the defn of h

Theorem~\ref{thm:generators} finds application in models where the probabilities $p_{\alpha, \beta}$ have the entire sample space $\mathbb G = \N^{\binom{n}{2}}$ as their support. In such scenarios, Theorem~\ref{thm:generators} enables the construction of a goodness-of-fit test for the exponential model by following \cite[Algorithm 1]{karwa2016exact}, augmented with an additional Metropolis-Hastings step in line 4 of the algorithm: given $m\in \mathcal M_{n,z}$ and a current graph $\gamma\in \fiberDC(d;c)$ we transition to a new graph $\gamma'=\gamma+m$ with probability $\min\big\{1, \frac{h(\gamma')}{h(\gamma)}\big\}$ whenever $\gamma'\in\fiberDC(d;c)$; otherwise, we remain at the current graph $\gamma$.

As an example, we could conduct the goodness-of-fit test when $\gamma_{uv}$ is modeled using a Poisson distribution with parameter $\beta_u+\beta_v+\alpha_{z(u)z(v)}$ and $(\alpha, \beta)\in\Theta = \R_{>0}^{\binom{k+1}{2}}\times \R_{>0}^{n}$.

\section{Restriction to simple graphs}\label{section:simple_graphs}

Given the result of Theorem~\ref{thm:generators} we know that $\mathcal M_{n,z}=\{[uvu'v']: z(u)=z(u') \text{ or }z(v)=z(v')\}$  can be used to find $c$-degree preserving paths between any two multigraphs with the same $c$-degree sequence. Hence, a natural question arises as to whether an analogous result holds when strictly considering simple graphs. In other words, is the graph $\sfiberDC(d;c)_{\mathcal M_{n,z}}$ connected for every color-graphic degree sequence $(d;c)\in \N^{n+\binom{k+1}{2}}$? In this section, we provide a negative answer to this question and discuss some open problems derived from our results.

\begin{prop}\label{prop:large_indispensable}
    For every integer $k\geq 3$ there exists a $k$-coloring $z$ of $[n]$ with $n=2k$, and a $c$-degree sequence $(d_k;c_k)\in \N^{n+\binom{k+1}{2}}$ such that $\widetilde{\mathcal F}_{\scriptsize{\DC}_{\!n,z}}(d_k;c_k)=\{\gamma, \gamma'\}$. Furthermore, $||\gamma-\gamma'||_1=2k$.    
\end{prop}

\begin{proof}
Let $k\geq 3$, be an integer and $z$ be the $k$-coloring of $[2k]$ such that $z(u)\equiv u\pmod k$ for every $u\in [2k]$. Let $3_k$ and $1_k$ be the vectors of size $k$ with all 3's and all 1's, respectively. Consider $d_k=(3_k; 1_k)\in \N^{2k}$ and $c_k\in \N^{\binom{k+1}{2}}$ be such that for every $1\leq u\leq v\leq k$ 
\[
c_k(u,v)=
\begin{cases}
    2 & \text{if } u=v+1 \text{ and }v< k, \text{ or } (u,v)=(1,k)\\
    0 & \text{otherwise}.
\end{cases}
\]

In order to prove that the simple-graph fiber $\widetilde{\mathcal F}_{\scriptsize{\DC}_{\!2k,z}}(d_k;c_k)$ contains only two elements we start by making the following two claims for every $\gamma\in \widetilde{\mathcal F}_{\scriptsize{\DC}_{\!2k,z}}(d_k;c_k)$ which will be shown later in this proof. 
\begin{enumerate}[align=left, label=\textit{Claim \arabic*}.,  labelsep=.5em]
    \item \label{item:claim_1} If $\{u+k,v\}\in E(\gamma)$ for $u\in [k]$, $v\in [2k]$ with $v\equiv u +\ep\pmod{k}$ and $\ep\in \{1,-1\}$; then $\{u,w\}\in E(\gamma)$ for any $w\in[2k]$ with $w\equiv u-\ep \pmod{k}$. 
    % $\{u,u-\ep\}$ and $\{u,(u+n)-\ep\}$ are both edges in $E(g)$. 
    \item \label{item:claim_2} The set $\{k+1, k+2, \ldots, 2k\}$ is independent in $\gamma$.
\end{enumerate}
Assuming both claims, let $\gamma$ be a simple graph in $\widetilde{\mathcal F}_{\scriptsize{\DC}_{\!2k,z}}(d_k;c_k)$. Since $c_k(1,v)\neq 0$ if and only if $v\in \{2, k\}$, it follows from Claim 2 that either $\{1+k, 2\}$ or $\{1+k, k\}$ is an edge in $E(\gamma)$ but not both. We will prove that choosing one of the previous two edges determines $\gamma$ completely. Assume that $\{1+k,2\}$ is an edge in $E(\gamma)$. By Claim 1 we have that $\{1,k\}$ and $\{1,2k\}$ are both edges in $E(\gamma)$. Now, since $\{2k, 1\}\in E(\gamma)$ it follows from Claim 1 that $\{k,k-1\}$ and $\{k, 2k-1\}$ are both edges in $E(\gamma)$. Continuing with this process and repeatedly applying Claim 1 shows that for any $u\in [k]$ and $v\in [2k]$ such that $v\equiv u-1 \pmod{k}$, $\{u,v\}\in E(\gamma)$. Let $E\subset E(\gamma)$ be the set with all the edges of this form and let $\gamma_E$ be the subgraph of $\gamma$ generated by $E$. Then, $(\deg_{\gamma_E}(u))_{u\in [2k]}=(3_k; 1_k)$. This implies that $E=E(\gamma)$, which means $\gamma=\gamma_E$. 

An analogous argument shows that if $\{1+k,k\}$ is an edge in $\gamma$ (instead of $\{1+k, 2\}$) then the graph $\gamma$ is generated by the set of edges $E'=\{\{u,v\}:u\in [k], v\in [2k]\text{ and }v\equiv u+1 \pmod{k}\}$. This shows that the only two graphs in $\widetilde{\mathcal F}_{\scriptsize{\DC}_{\!2k,z}}(d_k;c_k)$ are $\gamma_E$ and $\gamma_{E'}$. Since $E \cap E'=\{\{u,v\}:u,v\in [k] \text{ and }v-u\equiv 1 \pmod{k}\}$, we conclude that $||g_E-g_{E'}||_1=|E|+|E'|-2|E\cap E'|=2k$. 

Now we prove claims 1 and 2. Suppose $\gamma\in \widetilde{\mathcal F}_{\scriptsize{\DC}_{\!2k,z}}(d_k;c_k)$ and let $\{u+k,v\}\in E(\gamma)$ where $u\in [k]$, $v\in [2k]$ are such that $v\equiv u +\ep\pmod{k}$ with $\ep\in \{1,-1\}$. Let $w, w'\in [k]$ such that $w\equiv u+\ep \pmod{k}$ and $w'\equiv u-\ep \pmod{k}$. Since $c_k(u,w')=2$, $z^{-1}(w')=\{w', w'+k\}$ and $\deg_\gamma(u+k)=1$, it follows that $\{u, w'+k\}$ and $\{u, w'\}$ are both edges in $E(\gamma)$. This proves Claim 1. To prove Claim 2 let $\gamma\in \fiberDC(d_k;c_k)$ and suppose that $\{k+1, k+2, \ldots, 2k\}$ is not independent. Without losing generality assume that $\{k+1,k+2\}$ is an edge in $E(\gamma)$. Claim 1 then implies that $\{1, k\}$ and $\{1, 2k\}$ are both edges in $E(\gamma)$. Following an argument analogous to the proof of Claim 1 we can see that for every $u\in [k]$, $\{u, v\}\in E(\gamma)$ for any $v\in [2k]$ such that $v\equiv u-1 \pmod{k}$. In particular, this means that $\{3, 2+k\}\in E(\gamma)$ which would imply that $\deg_\gamma(2+k)\geq 2$. By the definition of $d_k$, this is a contradiction. Therefore, the set $\{k+1, k+2, \ldots, 2k\}$ must be independent.

\end{proof}

\begin{example}
    The simple graphs $\gamma_1,\gamma_2$ in Figure \ref{fig:indispensable_hexagons} represent the only two elements of the simple-graph fiber $\widetilde{\mathcal F}_{A_{12,z}}(d_6;c_6)$ where $z: [12] \to [6]$ is such that $z(u)\equiv u \pmod{6}$ for every $u\in [12]$ and $d_6,c_6$ are defined as in the proof of Proposition~\ref{prop:large_indispensable}. The only move (up to sign) that connects this fiber is $\gamma=\gamma_1-\gamma_2$.

    \begin{figure}[h]

    \definecolor{color1}{RGB}{214, 39, 40}     % Red
    \definecolor{color2}{rgb}{0.01, 0.75, 0.24}    % darkpastelgreen
    \definecolor{color3}{rgb}{255,0,255}   % Magenta
    \definecolor{color4}{rgb}{1.0, 0.49, 0.0}   % Amber
    \definecolor{color5}{rgb}{0.0, 0.5, 1.0}     % azure(colorwheel)
    \definecolor{color6}{rgb}{0.54, 0.17, 0.89}  % blue-violet
    \begin{minipage}{0.33\textwidth}
    \centering
    \begin{tikzpicture}[scale=1, every node/.style={minimum size=5mm, text centered}]
        % Draw the inner hexagon
        \foreach \i/\color in {1/color1, 2/color2, 3/color3, 4/color4, 5/color5, 6/color6}{
            \node[draw, circle, preaction={fill=\color, fill opacity=0.1}, draw=\color, inner sep=0pt] (n\i) at ({-(\i-1)*60}:1) {\i};
        }
        \draw[thick] (n1) -- (n2) -- (n3) -- (n4) -- (n5) -- (n6) -- (n1);

        % Draw the outer hexagon
        \foreach \i/\color in {7/color1, 8/color2, 9/color3, 10/color4, 11/color5, 12/color6} {
            \node[draw, circle, preaction={fill=\color, fill opacity=0.1}, draw=\color, inner sep=0pt] (n\i) at ({-(\i-1-6)*60}:1.8) {$\i$};
        }

        \draw (n1) -- (n12);
        \draw (n2) -- (n7);
        \draw (n3) -- (n8);
        \draw (n4) -- (n9);
        \draw (n5) -- (n10);
        \draw (n6) -- (n11);
    \end{tikzpicture}
    \end{minipage}%
    \begin{minipage}{0.33\textwidth}
    \centering
    \begin{tikzpicture}[scale=1, every node/.style={minimum size=5mm, text centered}]
        % Draw the inner hexagon
        \foreach \i/\color in {1/color1, 2/color2, 3/color3, 4/color4, 5/color5, 6/color6}{
            \node[draw, circle, preaction={fill=\color, fill opacity=0.1}, draw=\color, inner sep=0pt] (n\i) at ({-(\i-1)*60}:1) {\i};
        }
        \draw[thick] (n1) -- (n2) -- (n3) -- (n4) -- (n5) -- (n6) -- (n1);

        % Draw the outer hexagon
        \foreach \i/\color in {7/color1, 8/color2, 9/color3, 10/color4, 11/color5, 12/color6} {
            \node[draw, circle, preaction={fill=\color, fill opacity=0.1}, draw=\color, inner sep=0pt] (n\i) at ({-(\i-1-6)*60}:1.8) {$\i$};
        }

        \draw (n1) -- (n8);
        \draw (n2) -- (n9);
        \draw (n3) -- (n10);
        \draw (n4) -- (n11);
        \draw (n5) -- (n12);
        \draw (n6) -- (n7);
    \end{tikzpicture}
    \end{minipage}%
    \begin{minipage}{0.33\textwidth}
    \centering
    \begin{tikzpicture}[scale=1, every node/.style={minimum size=5mm, text centered}]
        % Draw the inner hexagon
        \foreach \i/\color in {1/color1, 2/color2, 3/color3, 4/color4, 5/color5, 6/color6}{
            \node[draw, circle, preaction={fill=\color, fill opacity=0.1}, draw=\color, inner sep=0pt] (n\i) at ({-(\i-1)*60}:1) {\i};
        }

        % Draw the outer hexagon
        \foreach \i/\color in {7/color1, 8/color2, 9/color3, 10/color4, 11/color5, 12/color6} {
            \node[draw, circle, preaction={fill=\color, fill opacity=0.1}, draw=\color, inner sep=0pt] (n\i) at ({-(\i-1-6)*60}:1.8) {$\i$};
        }

        \draw (n1) -- (n12);
        \draw (n2) -- (n7);
        \draw (n3) -- (n8);
        \draw (n4) -- (n9);
        \draw (n5) -- (n10);
        \draw (n6) -- (n11);
        
        \draw[dashed] (n1) -- (n8);
        \draw[dashed] (n2) -- (n9);
        \draw[dashed] (n3) -- (n10);
        \draw[dashed] (n4) -- (n11);
        \draw[dashed] (n5) -- (n12);
        \draw[dashed] (n6) -- (n7);
    \end{tikzpicture}
    \end{minipage}
    \caption{Simple graphs $\gamma_1$ and $\gamma_2$ on the left and center. Markov basis move $\gamma_1-\gamma_2$ on the right.}
    \label{fig:indispensable_hexagons}
    \end{figure}
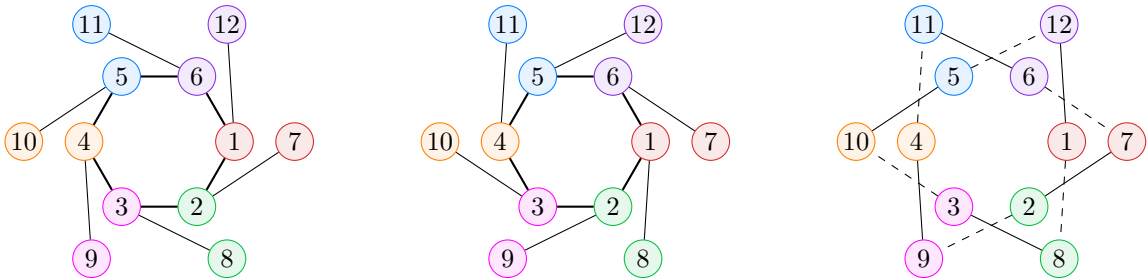
\end{example}

As an immediate consequence of Proposition~\ref{prop:large_indispensable} we have the following two results.

\begin{corollary}\label{cor:two_kappa}
Let $n, k \in \mathbb{Z}_+$ with $k \geq 3$, and let $z$ be a $k$-coloring of $[n]$. 
Then any simple-graph Markov basis for $\DC_{\!n,z}$ contains an element with $1$-norm equal to $2\kappa$, 
where $\kappa = \#\{\, i \in [k] : |z^{-1}(i)| \geq 2 \,\}$.
\end{corollary}

\begin{proof}
Let $\{\, i \in [k] : |z^{-1}(i)| \geq 2 \,\} = \{i_1, i_2, \ldots, i_\kappa\}$, 
and for each $j \in [\kappa]$ choose distinct vertices 
$u_j, u_{\kappa + j} \in z^{-1}(i_j)$. 
Define $z_0 : [2\kappa] \to [\kappa]$ by $z_0(u) \equiv u \pmod{\kappa}$ for every $u \in [2\kappa]$. 
As a consequence of the proof of Proposition~\ref{prop:large_indispensable}, there exists a $c$-degree sequence 
$(d^{(0)}; c^{(0)}) \in \mathbb{N}^{2\kappa + \binom{\kappa + 1}{2}}$ such that 
\[
\widetilde{\mathcal{F}}_{\scriptsize{\DC}_{\!2\kappa, z_0}}(d^{(0)}; c^{(0)}) = \{ \gamma, \gamma' \}
\quad \text{and} \quad 
\|\gamma - \gamma'\|_1 = 2\kappa.
\]

Now define $(d; c) \in \mathbb{N}^{n + \binom{k + 1}{2}}$ by
\[
d_u = 
\begin{cases}
0, & \text{if } u \notin \{u_1, u_2, \ldots, u_{2\kappa}\},\\[4pt]
d^{(0)}_j, & \text{if } u = u_j \text{ for } j \in [2\kappa],
\end{cases}
\]
and
\[
c(i, i') = 
\begin{cases}
0, & \text{if } \{i, i'\} \not\subseteq \{i_1, \ldots, i_\kappa\},\\[4pt]
c^{(0)}(l, m), & \text{if } i = i_l,\, i' = i_m.
\end{cases}
\]

Let $\psi : \mathbb{R}^{2\kappa + \binom{\kappa + 1}{2}} \longrightarrow \mathbb{R}^{n + \binom{k + 1}{2}}$
be the map defined as follows. For 
\[
a = (a_1, \ldots, a_{2\kappa};\, (a(l, m) : 1 \leq l \leq m \leq \kappa)) \in \mathbb{R}^{2\kappa + \binom{\kappa + 1}{2}},
\]
set $\psi(a) = \hat{a} = ((\hat{a}_u)_{u=1}^n;\, (\hat{a}(i, i') : 1 \leq i \leq i' \leq k))$, where
\[
\hat{a}_u =
\begin{cases}
0, & \text{if } u \notin \{u_1, u_2, \ldots, u_{2\kappa}\},\\[4pt]
a_j, & \text{if } u = u_j \text{ for } j \in [2\kappa],
\end{cases}
\]
and
\[
\hat{a}(i, i') =
\begin{cases}
0, & \text{if } \{i, i'\} \not\subseteq \{i_1, \ldots, i_\kappa\},\\[4pt]
a(l, m), & \text{if } i = i_l,\, i' = i_m.
\end{cases}
\]

Since $\psi$ is an isometric embedding, it follows that
\[
\widetilde{\mathcal{F}}_{\scriptsize{\DC}_{\!n, z}}(d; c) 
\cong 
\widetilde{\mathcal{F}}_{\scriptsize{\DC}_{\!2\kappa, z_0}}(d^{(0)}; c^{(0)}).
\]
Moreover,
\[
\widetilde{\mathcal{F}}_{\scriptsize{\DC}_{\!n, z}}(d; c) = \{ \hat{\gamma}, \hat{\gamma}' \}
\quad \text{and} \quad 
\|\hat{\gamma} - \hat{\gamma}'\|_1 = 2\kappa.
\]
Hence, there exists an element in the simple-graph Markov basis for $\DC_{\!n, z}$ with $1$-norm equal to $2\kappa$, as claimed.
\end{proof}

\thmtwo*

\begin{proof}
Let $k \in \Z$ with $k > C/2$, and set $n = 2k$. 
Define the $k$-coloring $z : [n] \to [k]$ by 
\[
z(u) \equiv u \pmod{k} \quad \text{for every } u \in [n].
\]
Notice that in this case, $\kappa = \#\{i \in [k] : |z^{-1}(i)| \ge 2\} = k.$
By Corollary~\ref{cor:two_kappa}, any simple-graph Markov basis for $\DC_{\!n,z}$ contains an element whose $1$-norm equals $2\kappa = 2k > C$.
\end{proof}

As it has been previously mentioned, when the $k$-coloring $z$ is constant, $4$-edge switches are enough to connect the space of simple graphs with a fixed degree sequence $d\in \N^n$ for any $d$. In contrast, Theorem \ref{thm:no_universal_constant} shows that when we do not impose any constraints on the coloring function $z$, we cannot guarantee the existence of a constant $C$ such that the set of $C$-edge switches induces connectivity on the space of simple graphs with fixed $c$-sequence $(d;c)$ for any color-graphical sequence $(d;c)\in \N^{n+\binom{k+1}{2}}$.

\begin{question}\label{question:1-norm_bound} Given $k\in \Z_+$, is there a constant $C_k$ such that for any $n\in\Z_+$ and any $k$-coloring $z$ of $[n]$, there exists a simple-graph Markov basis $\mathcal B$ for $\DC_{\!n,z}$ such that $\max_{\gamma\in \mathcal B}||\gamma||_1\leq C_k$? If so, what is the minimum $C_k$ satisfying this condition?
\end{question}

For $k=1$, $C_1=4$ is the minimum constant that satisfies the conditions in Question~\ref{question:1-norm_bound}. For $k=2$, Example~\ref{example:indispensable_squares} below shows that if $C_2$ exists it would neeed to be at least 8. 

\begin{example}\label{example:indispensable_squares} Consider the graphs $\gamma_1, \gamma_2$ from Example~\ref{example:connecting_multigraphs} which are the only two in their respective simple-graph fiber. In Figure~\ref{fig:indispensable_squares} we illustrate the move needed (up to sign) in the simple-graph Markov basis of $\DC_{8,z}$ in order to connect $\gamma_1$ and $\gamma_2$.

\begin{figure}[h]
\definecolor{Rcolor}{RGB}{214, 39, 40}    % Red vertices
\definecolor{Gcolor}{RGB}{44, 160, 44}    % Green vertices
\definecolor{Bcolor}{rgb}{0.0, 0.5, 1.0}  % Blue vertices
\begin{minipage}{0.33\textwidth}
\centering
\begin{tikzpicture}[scale=1, every node/.style={minimum size=5mm, text centered}]
    % Draw the inner square
    \foreach \i/\color/\label in {1/Bcolor/1, 2/Bcolor/2, 3/Rcolor/3, 4/Rcolor/4}{
        \node[draw, circle, preaction={fill=\color, fill opacity=0.1}, draw=\color, inner sep=0pt] (n\i) at ({-(\i-1)*360/4+45}:1) {\label};
    }

    % Draw the outer square
    \foreach \i/\color/\label in {5/Bcolor/8, 6/Bcolor/5, 7/Rcolor/6, 8/Rcolor/7} {
        \node[draw, circle, preaction={fill=\color, fill opacity=0.1}, draw=\color, inner sep=0pt] (n\i) at ({-(\i-1-4)*360/4+45}:2) {$\i$};
    }

    \draw (n1) -- (n4);
    \draw (n2) -- (n3);
    \draw (n2) -- (n4);
    \draw (n2) to[out=50,in=260] (n5);
    \draw (n2) -- (n6);
    \draw (n2) to[out=220,in=10] (n7);
    \draw (n2) to[out=105,in=-15] (n8);
    \draw (n4) to[out=40,in=190] (n5);
    \draw (n4) to[out=-75,in=165] (n6);
    \draw (n4) to[out=230,in=80] (n7);
    \draw (n4) -- (n8);
    \draw (n5) -- (n6);
    % \draw (n5) .. controls (2.3,-2.3) .. (n7);
    \draw (n5) .. controls (2.35, -2.3) and (2.3, -2.35) .. (n7);
    \draw (n7) -- (n8);
    
\end{tikzpicture}
\end{minipage}%
\begin{minipage}{0.33\textwidth}
\centering
\begin{tikzpicture}[scale=1, every node/.style={minimum size=5mm, text centered}]
    % Draw the inner square
    \foreach \i/\color/\label in {1/Bcolor/1, 2/Bcolor/2, 3/Rcolor/3, 4/Rcolor/4}{
        \node[draw, circle, preaction={fill=\color, fill opacity=0.1}, draw=\color, inner sep=0pt] (n\i) at ({-(\i-1)*360/4+45}:1) {\label};
    }

    % Draw the outer square
    \foreach \i/\color/\label in {5/Bcolor/8, 6/Bcolor/5, 7/Rcolor/6, 8/Rcolor/7} {
        \node[draw, circle, preaction={fill=\color, fill opacity=0.1}, draw=\color, inner sep=0pt] (n\i) at ({-(\i-1-4)*360/4+45}:2) {$\i$};
    }

    \draw (n5) -- (n8);
    \draw (n6) -- (n7);
    \draw (n2) -- (n4);
    \draw (n2) to[out=50,in=260] (n5);
    \draw (n2) -- (n6);
    \draw (n2) to[out=220,in=10] (n7);
    \draw (n2) to[out=105,in=-15] (n8);
    \draw (n4) to[out=40,in=190] (n5);
    \draw (n4) to[out=-75,in=165] (n6);
    \draw (n4) to[out=230,in=80] (n7);
    \draw (n4) -- (n8);
    \draw (n1) -- (n2);
    % \draw (n5) .. controls (2.3,-2.3) .. (n7);
    \draw (n5) .. controls (2.35, -2.3) and (2.3, -2.35) .. (n7);
    \draw (n3) -- (n4);
    
\end{tikzpicture}
\end{minipage}%
\begin{minipage}{0.33\textwidth}
\centering
\begin{tikzpicture}[scale=1, every node/.style={minimum size=5mm, text centered}]
    % Draw the inner square
    \foreach \i/\color/\label in {1/Bcolor/1, 2/Bcolor/2, 3/Rcolor/3, 4/Rcolor/4}{
        \node[draw, circle, preaction={fill=\color, fill opacity=0.1}, draw=\color, inner sep=0pt] (n\i) at ({-(\i-1)*360/4+45}:1) {\label};
    }

    % Draw the outer square
    \foreach \i/\color/\label in {5/Bcolor/8, 6/Bcolor/5, 7/Rcolor/6, 8/Rcolor/7} {
        \node[draw, circle, preaction={fill=\color, fill opacity=0.1}, draw=\color, inner sep=0pt] (n\i) at ({-(\i-1-4)*360/4+45}:2) {$\i$};
    }

    \draw (n1) -- (n4);
    \draw (n2) -- (n3);
    \draw (n7) -- (n8);
    \draw (n5) -- (n6);
    \draw[dashed] (n5) -- (n8);
    \draw[dashed] (n7) -- (n6);
    \draw[dashed] (n3) -- (n4);
    \draw[dashed] (n1) -- (n2);

\end{tikzpicture}    
\end{minipage}
\caption{Simple graphs $\gamma_1$ and $\gamma_2$ on the left and center. Markov basis move $\gamma_1-\gamma_2$ on the right.}
\label{fig:indispensable_squares}
\end{figure}
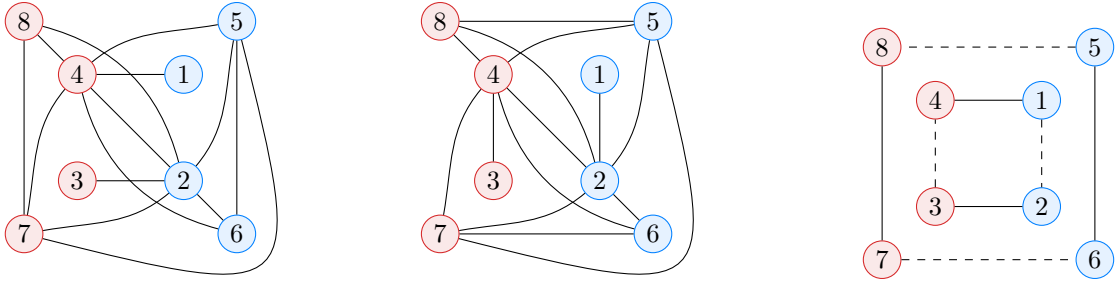
\end{example} 

\begin{conj}\label{conj:1-norm_bound_k2}
For $k=2$, $C_2$ exists and $C_2=8$ is the minimum constant satisfying the condition in Question~\ref{question:1-norm_bound}.   
\end{conj}

\section{A quadratic Gr\"obner basis}\label{sec:grobner}

The aim of this section is to show that $\{x^{\gamma^+}-x^{\gamma^-}: \gamma\in \mathcal M_{n,z}\}$ is in fact a Gr\"obner basis for $\IDC$ with respect to a monomial order defined below. When the $k$-coloring $z$ is constant, the statement follows directly from \cite[Theorem 2.1]{DeLoera1995hypersimplex}. As a matter of fact we will use this result, stated in Proposition~\ref{prop:jesus-thm}, as the motivation to prove Theorem~\ref{thm:Grobner_basis}.

To prove the main result of this section, we begin by introducing the monomial order $\succ$ as follows. 
We identify the set $[n]$ with the vertices of a complete graph $K_n$ embedded in the plane so that the vertices form a regular $n$-gon labeled clockwise from $1$ to $n$. 
We define the \emph{weight} of the variable $x_{uv}$ as the number of edges of $K_n$ that do not meet the edge $uv$. 
For instance, if $n = 5$, then the variables $x_{12}, x_{23}, x_{34}, x_{45}, x_{15}$ have weight $3$, while the variables $x_{13}, x_{24}, x_{35}, x_{14}, x_{25}$ have weight $1$. 
In general, the weight of a monomial $x^\alpha := \prod_{uv} x_{uv}^{\alpha_{uv}}$ is defined as the sum of the weights of the variables $x_{uv}$ appearing in $x^\alpha$, counted with multiplicity. 

\begin{lemma}\label{lemma:in-monomial-of-diff}
Let $\succ$ be any monomial order on $K[x_{uv} : 1 \leq u < v \leq n]$ that refines the partial order on monomials induced by the weights defined above. 
Given any pair of non-intersecting edges $uv, u'v'$ of $K_n$ such that $uv'$ and $vu'$ intersect, we have
\[
\text{in}_\succ (x_{uv}x_{u'v'} - x_{uv'}x_{u'v}) = x_{uv}x_{u'v'}.
\]
\end{lemma}

\begin{proof}
We proceed by induction on $n$. 
For $n = 4$, the statement is immediate. 
Assume that the claim holds for $n$, and let $\succ$ be a monomial order on $K[x_{uv} : 1 \leq u < v \leq n+1]$ refining the partial order on monomials induced by the weights. 
Let $uv, u'v'$ be a pair of edges in $K_{n+1}$ such that $uv'$ and $vu'$ intersect. 
Since $n > 4$, there exists a vertex $w \in [n+1] \setminus \{u, v, u', v'\}$.

For any pair of vertices $u_0, v_0 \neq w$, let $\mathcal{W}_w(u_0, v_0)$ denote the number of edges in $K_n$ that do not meet the edge $u_0v_0$ and have $w$ as an endpoint. 
Similarly, let $\mathcal{W}_{\hat{w}}(u_0, v_0)$ denote the number of edges in $K_n$ that do not meet $u_0v_0$ and do not have $w$ as an endpoint. Thus, the weight of the variable $x_{u_0v_0}$ is equal to $\mathcal{W}_w(u_0, v_0) + \mathcal{W}_{\hat{w}}(u_0, v_0)$.
By the induction hypothesis, we have
\[
\mathcal{W}_{\hat{w}}(uv) + \mathcal{W}_{\hat{w}}(u'v') 
> 
\mathcal{W}_{\hat{w}}(uv') + \mathcal{W}_{\hat{w}}(u'v).
\]
It therefore suffices to show that
\[
\mathcal{W}_w(uv) + \mathcal{W}_w(u'v') 
\geq 
\mathcal{W}_w(uv') + \mathcal{W}_w(u'v),
\]
which will imply the desired inequality for $n+1$.

For each pair $\{u_0, v_0\}$ in $\{\{u, v\}, \{u, u'\}, \{u', v'\}, \{v, v'\}\},$ there exist two circular paths between $u_0$ and $v_0$ with respect to the clockwise embedding of $K_{n+1}$ in the plane: 
one that contains all the vertices in $\{u, v, u', v'\}$, and another whose only vertices in $\{u, v, u', v'\}$ are $u_0$ and $v_0$. 
Denote the latter by $P_{u_0v_0}$, and let $\ell_{u_0v_0}$ be the number of vertices in this path. 
Then
\[
\ell_{uv} + \ell_{uu'} + \ell_{u'v'} + \ell_{vv'} = n + 5.
\]
Moreover, since $w \in [n+1] \setminus \{u, v, u', v'\}$, the vertex $w$ belongs to exactly one of the sets $P_{uv}$, $P_{uu'}$, $P_{u'v'}$, or $P_{vv'}$. 
The corresponding values of $\mathcal{W}_w(u_0, v_0)$ in each case are summarized in the following table.

\begin{table}[h!]
\centering
\begin{tabular}{c|c|c|c|c}
 & $\mathcal{W}_w(uv)$ & $\mathcal{W}_w(u'v')$ & $\mathcal{W}_w(uv')$ & $\mathcal{W}_w(u'v)$ \\ \hline
$w \in P_{uv}$ & $\ell_{uv} - 3$ & $n - \ell_{u'v'}$ & $\ell_{uv} + \ell_{vv'} - 5$ & $\ell_{uv} + \ell_{uu'} - 5$ \\ \hline
$w \in P_{uu'}$ & $n - \ell_{uv}$ & $n - \ell_{u'v'}$ & $\ell_{uu'} + \ell_{u'v'} - 5$ & $\ell_{uv} + \ell_{uu'} - 5$ \\ \hline
$w \in P_{u'v'}$ & $n - \ell_{uv}$ & $\ell_{u'v'} - 3$ & $\ell_{uu'} + \ell_{u'v'} - 5$ & $\ell_{u'v'} + \ell_{vv'} - 5$ \\ \hline
$w \in P_{vv'}$ & $n - \ell_{uv}$ & $n - \ell_{u'v'}$ & $\ell_{uv} + \ell_{vv'} - 5$ & $\ell_{u'v'} + \ell_{vv'} - 5$
\end{tabular}
\end{table}

In each case, one verifies that 
\[
\mathcal{W}_w(uv) + \mathcal{W}_w(u'v') 
\geq 
\mathcal{W}_w(uv') + \mathcal{W}_w(u'v),
\]
using the identity $\ell_{uv} + \ell_{uu'} + \ell_{u'v'} + \ell_{vv'} = n + 5$. 
This completes the inductive step and hence the proof of the lemma.
\end{proof}

Given a monomial order $\succ$ that refines the partial order on monomials specified by the weights, Lemma~\ref{lemma:4cycles} and Lemma~\ref{lemma:in-monomial-of-diff} imply that 
\begin{align}\label{eqn:initial_monomials}
\{\text{in}_\succ(x^{\gamma^+}-x^{\gamma^-}):\gamma\in \mathcal M_{n,z}\} = \{ &\; x_{uv}x_{u'v'} :uv,u'v'\text{ do not intersect in the embedding of }K_n \notag\\ &\text{ in the plane and }\{z(u), z(v)\}\cap\{z(u'), z(v')\}\neq \emptyset \}.
\end{align}

\begin{proposition}[\cite{DeLoera1995hypersimplex}, Theorem 2.1]\label{prop:jesus-thm} The set of binomials $\{x^{\gamma^+}-x^{\gamma^-}: \gamma\in \ker_\Z D_n, ||\gamma||_1=4\}$ is a Gr\"obner basis for $I_{D_n}$ with respect to $\succ$.
\end{proposition}

\begin{proposition}\label{prop:grobner-monochromatic} For any monomial walk $\gamma\in \ker_\Z\!\DC_{\!n,z}$ there exists a pair of non-intersecting edges $uv, u'v'$ in the embedding of $K_n$ in the plane such that $x_{uv}x_{u'v'}$ divides either $x^{\gamma^+}$ or $x^{\gamma^-}$.
\end{proposition}

\begin{proof}
Let $\gamma\in \ker_\Z \!\DC_{\!n,z}$ and let $\mathcal M_n=\{\omega\in \ker_\Z D_n: ||\omega||_1=4\}$. Similarly to \eqref{eqn:initial_monomials}, from the definition of $\succ$ it follows that
\begin{align*}
\{\text{in}_\succ(x^{\omega^+}-x^{\omega^-}):\omega\in \mathcal M_{n}\} = \{ &\; x_{uv}x_{u'v'} :uv,u'v'\text{ do not intersect in the embedding of }K_n \notag\\ &\text{ in the plane}\}.
\end{align*}
As a consequence of the containment $\ker_\Z \!\DC_{\!n,z}\subseteq \ker_\Z D_n$, it follows that $x^{\gamma^+}-x^{\gamma^-}\in I_{D_n}$. Furthermore, Proposition~\ref{prop:jesus-thm} implies that $\text{in}_\succ(I_{D_n})=\langle\{\text{in}_\succ(x^{\omega^+}-x^{\omega^-}):\omega \in \mathcal M_n\}\rangle$. Hence, there exists $\omega\in \mathcal M_n$ such that $\text{in}_\succ(x^{\omega^+}-x^{\omega^-})=x_{uv}x_{u'v'}$ divides $\text{in}_\succ(x^{\gamma^+}-x^{\gamma^-})$. Since $uv$ and $u'v'$ don't intersect in the embedding of $K_n$ in the plane, the result follows.   
\end{proof}
 
To establish the proof for Theorem~\ref{thm:Grobner_basis}, we will introduce the following notation, lemmas and propositions. 

Let $z:[n]\to [k]$ and let $q, q'\in [k]$. We define the $k$-coloring $z_{q}^{q'}:[n]\to [k]$ as 
\begin{equation}\label{eqn:coloring_restriction}
z_{q}^{q'}(i):=
\begin{cases}
q & \text{if }z(i)=q',\\
z(i) & \text{otherwise.}
\end{cases} 
\end{equation}

In other words, the $k$-coloring $z_q^{q'}$ is obtained from the $k$-coloring $z$ by re-coloring all the $q'$-th colored vertices with the $q$-th color. Then, we have the following.

\begin{lemma}\label{lemma:restrict-coloring} For any $k$-coloring $z$ of $[n]$ and $q, q'\in [k]$ we have $\ker_\Z\!\DC_{\!n,z}\subseteq \ker_\Z A_{n,z_q^{q'}}$.
\end{lemma}

\begin{proof} Let $\gamma\in \ker_\Z\!\DC_{\!n,z}$ and $q, q'\in [k]$. $\gamma$ satisfies the degree-condition by assumption. Hence, all we need to prove is that $\gamma$ satisfies the color-balance condition with respect to $z_{q}^{q'}$. To do so, notice that for every $i,j$ different from $q'$,

\[
c_\gamma^{\pm}(z_q^{q'},i,j)=
\begin{cases}
c^{\pm}_{\gamma}(z,i,j) &\text{ if } i,j\neq q,\\
c^{\pm}_{\gamma}(z,i,j)+c^{\pm}_{\gamma}(z,q',j) &\text{ if } i=q, j\neq q,\\
c^{\pm}_{\gamma}(z,q,q)+c^{\pm}_{\gamma}(z,q',q')+c^{\pm}_{\gamma}(z,q,q') &\text{ if } i=j=q,\\
0 & \text{ if } i=q' \text{ or } j=q'.
\end{cases}
\]
\end{proof}

Now let $\gamma\in \Z^{\binom{n}{2}}$, $w\in [n]$, and $z:[n]\to[k]$ with $z(w)=q$. We define the \emph{contraction of $\gamma$ with respect to $w$ and $z$} as the vector $\sigma_w(\gamma)\in \Z^{\binom{n}{2}}$ such that for every distinct $u,v\in [n]$,

\begin{equation}\label{def:contraction}
\sigma_w(\gamma)_{uv}:=
\begin{cases}
\gamma_{uv}, &\text{if } z(u)\neq q\text{ and }z(v)\neq q,\\
\sum\limits_{u'\in z^{-1}(q)}\gamma_{u'v}, & \text{if } u=w \text{ and } z(v)\neq q,\\
\sum\limits_{v'\in z^{-1}(q)}\gamma_{uv'}, & \text{if } v=w  \text{ and } z(u)\neq q,\\
0, &\text{otherwise }.
\end{cases}
\end{equation}

We call $\sigma_w(\gamma)$ simply a contraction when $w$ and $z$ are clear from the context.

\begin{rmk}\label{rmk:zero_coordinates} Notice that whenever $u$ or $v$ belong to the set $z^{-1}(q)\minus\{w\}$ we have $\sigma_w(\gamma)_{uv}=0$. In other words, $z^{-1}(q)\minus\{w\}$ is an isolated set of vertices in $\sigma_w(\gamma)$ when regarded as a graph. This implies that for any $v\in [n]$, $S\subseteq [n]\minus\{v\}$ and $S'\subseteq z^{-1}(q)\minus\{w\}$ $$\sum_{u\in S}\sigma_w(\gamma)_{uv}=\sum_{u\in S\minus S'}\sigma_w(\gamma)_{uv}=\sum_{u\in S\cup S'}\sigma_w(\gamma)_{uv}.$$ 
\end{rmk}

\begin{example}\label{example:reduction} Consider the monomial walk $\gamma$ from Example~\ref{example:monomial_walk}. The contraction $\sigma_1(\gamma)$ is shown in Figure \ref{fig:reduction}. Notice that the reduction $\sigma_3(\sigma_1(\gamma))$ returns a zero-vector, or in other words, an empty graph.

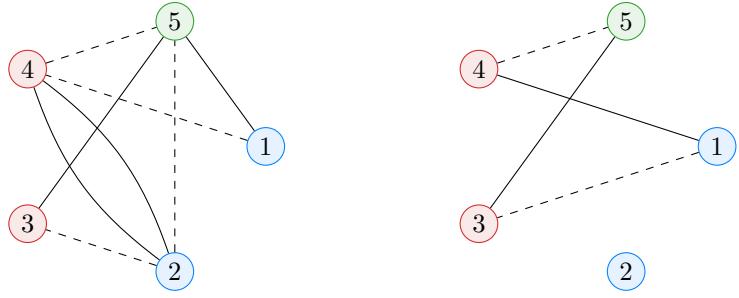
\begin{figure}[h]
\centering
\begin{tikzpicture}[scale=1, every node/.style={minimum size=5mm, text centered}]

   % Draw the circular graph
   \foreach \i/\col in {1/Bcolor, 2/Bcolor, 3/Rcolor, 4/Rcolor, 5/Gcolor} {
       \node[draw, circle, preaction={fill=\col, fill opacity=0.1}, draw=\col, inner sep=0pt] (n\i) at ({(-\i)*360/5+72}:1.75) {\footnotesize{\i}};
   }
   
   % Draw the edges
    \path
      (n2) edge[bend left=17] (n4)
      (n4) edge[bend left=17] (n2)
      (n1) edge (n5)
      (n3) edge (n5)
      (n2) edge[dashed] (n3)
      (n2) edge[dashed] (n5)
      (n4) edge[dashed] (n5)
      (n1) edge[dashed] (n4);

   % Duplicate the contents and shift to the right
   \begin{scope}[xshift=6cm]
       \foreach \i/\col in {1/Bcolor, 2/Bcolor, 3/Rcolor, 4/Rcolor, 5/Gcolor} {
           \node[draw, circle, preaction={fill=\col, fill opacity=0.1}, draw=\col, inner sep=0pt] (n\i) at ({(-\i)*360/5+72}:1.75) {\footnotesize{\i}};
       }
       \path
          (n1) edge (n4)
          (n3) edge (n5)
          (n1) edge[dashed] (n3)
          (n4) edge[dashed] (n5);
   \end{scope}
\end{tikzpicture}
\caption{Monomial walk $\gamma=[1,5,2,4,5,3,2,4]$ on the left and its contraction $\sigma_1(\gamma)=[1,4,5,3]$ on the right.} 
\label{fig:reduction}
\end{figure}

\end{example}

\begin{lemma} For any $k$-coloring $z$ of $[n]$, any monomial walk $\gamma\in \ker_\Z\!\DC_{\!n,z}$ and $w\in [n]$, $\sigma_w(\gamma)$ is also a monomial walk. In other words, the map $\sigma_w$ satisfies $\sigma_w(\ker_\Z\!\DC_{\!n,z})\subseteq \ker_\Z\!\DC_{\!n,z}$. 
\end{lemma}

\begin{proof}
Let $\gamma\in \ker_\Z\!\DC_{\!n,z}$ and $w\in [n]$ with $q:=z(w)$. We will show the following:

\begin{enumerate}
    \item $\sigma_w(\gamma)$ satisfies the degree-balance condition.

    Let $v\in [n]$. If $v\in z^{-1}(q)\minus\{w\}$, Remark~\ref{rmk:zero_coordinates} implies that $\deg^\pm_{\sigma_w(\gamma)}(v)=0$, so let us assume $v\notin z^{-1}(q)\minus\{w\}$ and consider the following two cases: 

    \begin{enumerate}[label=(\roman*)]
        \item $v\neq w$. In this case, by properly arranging summation indices and using Remark~\ref{rmk:zero_coordinates}, one can see that 
        {\allowdisplaybreaks
        \begin{align*}
        \deg_{\sigma_w(\gamma)}^+(v)-\deg_{\sigma_w(\gamma)}^-(v) 
        &=\sum_{u\neq v}\sigma_w(\gamma)_{uv} 
        = \sum\limits_{\substack{u\neq v \\ u\notin z^{-1}(q)}}\sigma_w(\gamma)_{uv}+\sum\limits_{\substack{u\neq v \\ u\in z^{-1}(q)}}\sigma_w(\gamma)_{uv}\\
        &= \sum\limits_{\substack{u\neq v \\ u\notin z^{-1}(q)}}\gamma_{uv}+\sigma_w(\gamma)_{wv}
        = \sum\limits_{\substack{u\neq v \\ u\notin z^{-1}(q)}}\gamma_{uv}+\sum\limits_{\substack{u'\in z^{-1}(q)}}\gamma_{u'v}\\
        &= \sum\limits_{\substack{u\neq v }}\gamma_{uv} 
        = \deg_\gamma^+(v)-\deg_\gamma^-(v)
        =0.
        % \\
        % &= \sum\limits_{\substack{u\neq v }}m_{uv} = \deg_m^+(v)-\deg_m^-(v)
        \end{align*}
        }   
        
        \item $v=w$. Similarly to the previous case, by strategically rearranging summation indices and using Remark~\ref{rmk:zero_coordinates}, we have
        {\allowdisplaybreaks
        \begin{align*}
        \deg_{\sigma_w(\gamma)}^+(v)-\deg_{\sigma_w(\gamma)}^-(v) 
        &=\sum_{u\neq w}\sigma_w(\gamma)_{uw} 
        = \sum\limits_{\substack{u\notin z^{-1}(q)}}\sigma_w(\gamma)_{uw}\\
        &= \sum\limits_{\substack{u\notin z^{-1}(q)}} \sum_{v'\in z^{-1}(q)}\gamma_{uv'}
        = \sum\limits_{\substack{i\in [k] \\ i\neq q}}\Big(\sum\limits_{\substack{u\in z^{-1}(i)}} \sum_{v'\in z^{-1}(q)}\gamma_{uv'}\Big)\\
        &= \sum\limits_{\substack{i\in [k] \\ i\neq q}} \left(c_\gamma^+(i,q)-c_\gamma^-(i,q)\right)=0.
        \end{align*}
        Where the last equality holds because $\gamma$ satisfies the color-balance conditions.
        } 
    \end{enumerate}

    \item $\sigma_w(\gamma)$ satisfies the color-balance condition.

    Let $1\leq i\leq j\leq k$. It follows from equations \eqref{eqn:pos-neg-color} and \eqref{def:contraction} that if $i,j\neq q$, $c^{\pm}_{\sigma_w(\gamma)}(i,j)=c^{\pm}_{\gamma}(i,j)$ and if $i=j=q$ then $c^{\pm}_{\sigma_w(\gamma)}(i,j)=0$. In either of these two cases we have $c^{+}_{\sigma_w(\gamma)}(i,j)=c^{-}_{\sigma_w(\gamma)}(i,j)$. Now, suppose that $i\neq q$ and $j=q$. In this case we have
    {\allowdisplaybreaks
    \begin{align*}
    c^+_{\sigma_w(\gamma)}(i,j)-c_{\sigma_w(\gamma)}^-(i,j) 
    &= \sum_{u\in z^{-1}(i)}\sigma_w(\gamma)_{uw} 
    = \sum_{u\in z^{-1}(i)} \sum_{v'\in z^{-1}(q)}\gamma_{uv'}\\
    &= c^+_{\gamma}(i,q)-c^-_{\gamma}(i,q)=0.
    \end{align*}
    } 
    The case $i=q, j\neq q$ is analogous to the latter case.
\end{enumerate}
\end{proof}

For the rest of this manuscript, we assume that the $k$-coloring $z$ is non-decreasing. This is, for all $1\leq u<v\leq n$, $z(u)\leq z(v)$. The assumption will be useful thanks to the following.

\begin{remark}\label{remark:non-intersecting}
Consider the embedding of $K_n$ in the plane and suppose the $k$-coloring $z:[n]\to [k]$ is non-decreasing. If $uv, u'v'$ are two non-intersecting edges with $z(v')\notin \{z(u), z(u'), z(v)\}$ then for any vertex $w$ such that $z(w)=z(v')$, we have that $uv, u'w$ are non-intersecting edges in the embedding of $K_n$. 
\end{remark}

\begin{lemma}\label{lemma:con-edges}
Let $z$ be a non-decreasing $k$-coloring of $[n]$ with $k\geq 2$ and $\gamma\in \ker_\Z\!\DC_{\!n,z}$ be a monomial walk. Let $w\in [n]$ and $uv$ an edge in $\sigma_w(\gamma)$. Then, 

\begin{enumerate}[label=(\roman*)]
    \item there exists a vertex $v_0\in [n]$ such that $uv_0$ is an edge in $E(\gamma)$ with same sign as $uv$ and $z(v_0)=z(v)$,
    \item if $u'v'$ is an edge in $E(\sigma_w(\gamma))$ with $u',v'\neq w$ such that $uv, u'v'$ do not intersect in the embedding of $K_n$ in the plane, then $uv_0$ and $u'v'$ do not intersect in the embedding of $K_n$ either.  
\end{enumerate}

\end{lemma}

\begin{proof}

\begin{enumerate}[label=(\roman*)]

\item Let $q=z(w)$ and $uv$ an edge of $\sigma_w(\gamma)$. By definition, we have that every vertex in $z^{-1}(q)\minus\{w\}$ is isolated in $\sigma_w(\gamma)$. Hence we have two options: either $u,v\neq w$ or $v=w$ (or $u=w$). If $u,v\neq w$, $uv$ is also an edge of $\gamma$ in which case we can set $v_0=v$. Now, assume that $v=w$ (the case $u=w$ is completely analogous).

Given that $uw$ is an edge in $\sigma_w(\gamma)$, we have that $\sigma_w(\gamma)_{uw}\neq 0$. Assume without loosing generatity that $\sigma_w(\gamma)_{uw}> 0$ (i.e., $uv$ is positive). By definition, $\sigma_w(\gamma)_{uw}=\sum_{v_{0}\in z^{-1}(q)}\gamma_{uv_{0}}$, which implies that $\gamma_{uv_{0}}>0$ for some $v_0\in z^{-1}(q)$. This means that $uv_0$ is a positive edge in $\gamma$ with $z(v_0)=q=z(w)$. We can apply a similar argument for when $\sigma_w(\gamma)_{uw}<0$.  

\item Given that $u',v'\neq w$, it follows from the construction of $\sigma_w(\gamma)$ that $u'v'$ is an edge of $\gamma$. If $v\neq w\Rightarrow uv_0=uv$ so the result follows trivially. If $v=w$, then $z(v)\notin \{z(u), z(u'), z(v')\}$ so Remark~\ref{remark:non-intersecting} implies that $uv_0$ and $u'v'$ are non-intersecting edges in the embedding of $K_n$ in the plane.

\end{enumerate}
\end{proof}

\begin{proposition}\label{prop:non-intersecting}
Let $z$ be a non-decreasing $k$-coloring of $[n]$. For any monomial walk $\gamma\in \ker_\Z\!\DC_{\!n,z}$ there is a pair of non-intersecting edges $uv, u'v'$ in the embedding of $K_n$ in the plane such that $z(u)=z(u')$ and $x_{uv}x_{u'v'}$ divides either $x^{\gamma^+}$ or $x^{\gamma^-}$. 

\end{proposition}

\begin{proposition}\label{prop:non-intersecting}
Let $z$ be a non-decreasing $k$-coloring of $[n]$. For any monomial walk $\gamma\in \ker_\Z\!\DC_{\!n,z}$ there is a pair of non-intersecting edges $uv, u'v'$ in the embedding of $K_n$ in the plane such that $z(u)=z(u')$ and $x_{uv}x_{u'v'}$ divides either $x^{\gamma^+}$ or $x^{\gamma^-}$. 

\end{proposition}

\begin{proof}

Let $\gamma\in \ker_\Z\!\DC_{\!n,z}$ be a monomial walk. Notice that a monomial $x_{uv}x_{u'v'}$ divides $x^{\gamma^+}$ or $x^{\gamma^-}$ if and only if $uv, u'v'$ is  pair of edges in $E(\gamma)$ with same sign. Let $z:[n]\to [k]$ be a $k$-coloring assumed to be non-decreasing and for convenience suppose $z([n])=\{1, \ldots, \kappa\}$ for some $\kappa \in \Z_+$. When $\kappa=1$ (i.e., $z$ is constant) the result follows from Proposition~\ref{prop:grobner-monochromatic}. Let us divide the proof for $\kappa\geq 2$ in the following two cases:

\begin{enumerate}[label=(\alph*)]
    \item There exists a monochromatic edge $xy \in E(\gamma)$ with $z(x) = z(y) = \iota$. 

For every $i \in [\kappa]$, let $u_i := \min\{u : z(u) = i\},$
and define 
\[
\widetilde{\gamma} := \sigma_{u_1}\bigl(\cdots \hat{\sigma}_{u_{\iota}}\bigl(\cdots \sigma_{u_\kappa}(\gamma) \cdots \bigr)\cdots \bigr),
\]
where $\hat{\sigma}_{u_{\iota}}$ indicates that the contraction with respect to $u_{\iota}$ is omitted. Let
\[
w \in \argmax_{v \in [n] \,:\, z(v) \neq \iota} \deg^+_{\widetilde{\gamma}}(v).
\]
Next, define the two-coloring $\widetilde{z} : [n] \to \{\iota, \tau\}$, where $\tau \in [k] \setminus \{\iota\}$, by 
\[
\widetilde{z}(u) = 
\begin{cases}
\iota, & \text{if } u \in z^{-1}(\iota),\\
\tau, & \text{otherwise.}
\end{cases}
\]
Let $\sigma_w(\widetilde{\gamma})$ denote the contraction of $\widetilde{\gamma}$ with respect to $w$ and $\widetilde{z}$. From Proposition~\ref{prop:grobner-monochromatic}, there exist edges $uv, u'v' \in E(\sigma_w(\widetilde{\gamma}))$ such that $uv$ and $u'v'$ do not intersect in the embedding of $K_n$ in the plane. Since $w$ is the only (potentially) non-isolated vertex of $\sigma_w(\widetilde{\gamma})$ with $\widetilde{z}(w) \neq \iota$, we may assume without loss of generality that $\widetilde{z}(u) = \widetilde{z}(v) = \widetilde{z}(u') = \iota$.  

By Lemma~\ref{lemma:con-edges}(i), there exists a vertex $v'_0 \in [n]$ such that $u'v'_0 \in E(\widetilde{\gamma})$ has the same sign as $u'v'$ and $\widetilde{z}(v'_0) = \widetilde{z}(v')$. Moreover, since $uv, u'v' \in E(\sigma_w(\widetilde{\gamma}))$ do not intersect in the embedding of $K_n$ and $u, v \neq w$, it follows from Lemma~\ref{lemma:con-edges}(ii) that $uv$ and $u'v'_0$ also form a pair of non-intersecting edges in the embedding of $K_n$ in the plane.  

Notice that $uv$ and $u'v'_0$ have the same sign. Then, after applying Lemma~\ref{lemma:con-edges} repeatedly to $\widetilde{\gamma}$, we obtain a vertex $v''_0 \in [n]$, which might or might not be the same as $v'_0$, such that the edges $uv, u'v''_0 \in E(\gamma)$ have the same sign and do not intersect in the embedding of $K_n$ in the plane. Since we assumed that $\widetilde{z}(u) = \widetilde{z}(u')$, and hence $z(u) = z(u')$, this completes the proof of Proposition~\ref{prop:non-intersecting} under the assumptions made for this case.

\begin{figure}[h]
\begin{minipage}{0.5\textwidth}
\centering
\begin{tikzpicture}[scale=.9, every node/.style={minimum size=4.5mm, text centered}]
   % Define colors
   \definecolor{colorA}{RGB}{214, 39, 40}    % Red vertices
   \definecolor{colorB}{RGB}{140, 86, 75}    % Brown vertices
   \definecolor{colorC}{RGB}{44, 160, 44}    % Green vertices
   \definecolor{colorD}{rgb}{1.0, 0.49, 0.0} % Orange vertices
   \definecolor{colorE}{rgb}{0.0, 0.5, 1.0}  % Blue vertices
   \definecolor{colorF}{rgb}{0.54, 0.17, 0.89}  % Purple vertices

   % Draw the circular graph
   \foreach \i/\color in {1/colorA, 2/colorA, 3/colorA, 4/colorA, 5/colorB, 6/colorC, 7/colorC, 8/colorC, 9/colorD, 10/colorD, 11/colorD, 12/colorD, 13/colorE, 14/colorE, 15/colorF, 16/colorF, 17/colorF} {
       \node[draw, circle, preaction={fill=\color, fill opacity=0.1}, draw=\color, inner sep=0pt] (n\i) at ({-(\i-1)*360/17}:2.5) {\footnotesize{\i}};
   }
   
   % Draw the edges
   \draw (n1) to[out=220,in=70] (n4);
   \draw[dashed, line width=1.5pt] (n1) edge (n7);
   \draw (n2) edge (n11);
   \draw[dashed] (n2) to[out=120,in=235] (n17);
   \draw (n3) edge (n8);
   \draw (n3) edge (n15);
   \draw[dashed, line width=1.5pt] (n3) edge (n4);
   \draw[dashed] (n3) edge (n10);
   \draw (n5) to[out=145,in=0] (n7);
   \draw[dashed] (n5) edge (n6);
   \draw (n6) edge (n7);
   \draw[dashed] (n7) edge (n8);
   \draw (n8) edge (n17);
   \draw[dashed] (n8) edge (n15);
   \draw (n9) edge (n14);
   \draw[dashed] (n9) to[out=60,in=285] (n11);
   \draw (n10) edge (n11);
   \draw[dashed] (n11) to[out=15,in=215] (n14);
   \draw (n12) to[out=335,in=190] (n16);
   \draw[dashed] (n12) to[out=320,in=190] (n17);
   \draw (n13) to[out=320,in=165] (n17);
   \draw[dashed] (n13) to[out=335,in=170] (n16);

   % Add the new node labeled with $m$
   \node[draw=none, text=gray] at (145:4) {$\gamma$};
\end{tikzpicture}
\end{minipage}%
\begin{minipage}{0.5\textwidth}
\centering
\begin{tikzpicture}[scale=.9, every node/.style={minimum size=4.5mm, text centered}]
   % Define colors
   \definecolor{colorA}{RGB}{214, 39, 40}      % Color for the first 4 vertices
   \definecolor{colorB}{RGB}{140, 86, 75}    % Color for the next 3 vertices
   \definecolor{colorC}{RGB}{44, 160, 44} % Color for the next 1 vertex
   \definecolor{colorD}{rgb}{1.0, 0.49, 0.0}   % Color for the next 4 vertices
   \definecolor{colorE}{rgb}{0.0, 0.5, 1.0}     % Color for the next 2 vertices
   \definecolor{colorF}{rgb}{0.54, 0.17, 0.89}  % Color for the last 3 vertices

   % Draw the circular graph
   \foreach \i/\color in {1/colorA, 2/colorA, 3/colorA, 4/colorA, 5/colorB, 6/colorC, 7/colorC, 8/colorC, 9/colorD, 10/colorD, 11/colorD, 12/colorD, 13/colorE, 14/colorE, 15/colorF, 16/colorF, 17/colorF} {
       \node[draw, circle, preaction={fill=\color, fill opacity=0.1}, draw=\color, inner sep=0pt] (n\i) at ({-(\i-1)*360/17}:2.5) {\footnotesize{\i}};
   }
   
   % Draw the edges
   \draw (n1) to[out=220,in=70] (n4);
   \draw[dashed, line width=1.5pt] (n1) to[out=200,in=50] (n6);
   \draw[dashed] (n2) edge (n15);
   \draw (n2) edge (n9);
   \draw (n3) edge (n15);
   \draw (n3) to[out=185,in=25] (n6);
   \draw[dashed, line width=1.5pt] (n3) edge (n4);
   \draw[dashed] (n3) edge (n9);
   \draw (n2) edge (n9);
   
   % Add the new node labeled with $m$
   \node[draw=none, text=gray] at (145:4) {$\widetilde \gamma$};
\end{tikzpicture}
\end{minipage}
\begin{center}
\begin{tikzpicture}[scale=.9, every node/.style={minimum size=4.5mm, text centered}]
   % Define colors
   \definecolor{colorA}{RGB}{214, 39, 40}      % Color for the first 4 vertices
   \definecolor{colorB}{RGB}{140, 86, 75}    % Color for the next 3 vertices
   \definecolor{colorC}{RGB}{44, 160, 44} % Color for the next 1 vertex
   \definecolor{colorD}{rgb}{1.0, 0.49, 0.0}   % Color for the next 4 vertices
   \definecolor{colorE}{rgb}{0.0, 0.5, 1.0}     % Color for the next 2 vertices
   \definecolor{colorF}{rgb}{0.54, 0.17, 0.89}  % Color for the last 3 vertices

   % Draw the circular graph
   \foreach \i/\color in {1/colorA, 2/colorA, 3/colorA, 4/colorA, 5/colorB, 6/colorC, 7/colorC, 8/colorC, 9/colorD, 10/colorD, 11/colorD, 12/colorD, 13/colorE, 14/colorE, 15/colorF, 16/colorF, 17/colorF} {
       \node[draw, circle, preaction={fill=\color, fill opacity=0.1}, draw=\color, inner sep=0pt] (n\i) at ({-(\i-1)*360/17}:2.5) {\footnotesize{\i}};
   }
   
   % Draw the edges
   \draw (n1) to[out=220,in=75] (n4);
   \draw (n3) edge (n9);
   \draw[dashed, line width=1.5pt] (n3) edge (n4);   
   \draw[dashed, line width=1.5pt] (n1) edge (n9);
   
   % Add the new node labeled with $m$
   \node[draw=none, text=gray] at (155:4) {$\sigma_w(\widetilde \gamma)$};
\end{tikzpicture}
\end{center}
\caption{Illustration of Proposition~\ref{prop:non-intersecting}, case (a): In this example we have the monochromatic edge $\{3,4\}$ in $E(\gamma)$. After building $\widetilde{\gamma}$, we are choosing $w=9$ and we are defining $\widetilde{z}$ such that $\widetilde{z}(u) = \tau$ for every $u \in [12] \setminus [4]$ and $\widetilde{z}(u) = \iota$ otherwise. Notice that $\{u,v\}=\{3,4\}, \{u',v'\}=\{1,9\}$ is a pair of edges in $E(\sigma_w(\widetilde{\gamma}))$ that have the same sign and do not intersect. After applying Lemma~\ref{lemma:con-edges} once to $\widetilde{\gamma}$ with respect to $\widetilde{z}$, we get $v_0'=6$. After applying the lemma once more we get $v_0''=7$. This way, we recover the pair of non-intersecting edges $\{u,v\}=\{3,4\}, \{u',v_0''\}=\{1,7\}$ with same sign in $E(\gamma)$, satisfying $z(u)=z(u')$.}
\end{figure}

\item There are no monochromatic edges in the monomial walk $\gamma$. 

We will prove Proposition~\ref{prop:non-intersecting} for this case by induction on $\kappa$. First, let us observe that Proposition~\ref{prop:grobner-monochromatic} guarantees the existence of a pair of edges $uv,u'v'\in E(\gamma)$, both with the same sign and such that do not intersect in the embedding of $K_n$ in the plane. Now consider the following cases:
    
\begin{enumerate}[label=(\roman*)]

    \item Let $\kappa\leq 3$. By the assumption at the beginning of (b) we have that $z(u)\neq z(v), z(u')\neq z(v')$. From the Pigeonhole principle either $z(u)=z(u')$, $z(u)=z(v')$, $z(v)=z(v')$ or $z(v)=z(u')$. This proves our base case. 

    \item Let $\kappa\geq 4$. Assume that Proposition \ref{prop:non-intersecting} holds for any instance of case (b) for which the size of the $k$-coloring's range is smaller than $\kappa$.  

    Let us first prove that there exists $\nu \in [n]$ such that $z(u), z(v)\notin \{\nu,\nu+1\}$ ($\nu+1=1$ if $\nu=n$): When $\kappa> 4$ this follows from the Pigeonhole principle. When $\kappa=4$ we can use that $z$ is non-decreasing to see that any pair of edges with vertex colors $1,3$ intersects with any pair of edges with vertex colors $2,4$. In such a case we can assume w.l.o.g that $z(u)=1, z(v)=2$ and set $i=3$.

    The existence of $\nu$ guarantees that for any $w\in z^{-1}(\{q,q'\})$ the contraction $\sigma_w(\gamma)$ is non-empty. Now, let $\nu'=\nu+1$, $q=z(\nu)$ and $q'=z(\nu')$. Let $z_q^{q'}$ be the $k$-coloring of $[n]$ as defined in \ref{eqn:coloring_restriction} and notice that by Lemma~\ref{lemma:restrict-coloring} $\gamma$ is a monomial walk with respect to $z_{q}^{q'}$, i.e., $\gamma\in \ker_\Z A_{n, z_{q}^{q'}}$. Now, pick $w\in[n]$ such that $z_q^{q'}(w)=q$ and notice that by previous observation $\sigma_w(\gamma)$ is non-empty. Then, either by case (a) above or by the inductive step applied to $\sigma_w(\gamma)$ with respect to $z_q^{q'}$, there exists a pair of edges $xy, x'y'\in E(\sigma_{w}(\gamma))$ such that $z_q^{q'}(x)=z_q^{q'}(x')$ and $xy, x'y'$ do not intersect in the embedding of $K_n$ in the plane. Since $w$ is the only (possibly) non-isolated vertex with $z_q^{q'}(w)=q$, it follows that $x,x'\neq w$ and by definition of $z_q^{q'}$ it follows that $z(x)=z_q^{q'}(x)=z_q^{q'}(x')=z(x')$. 

    Notice that at least one of the vertices $y,y'$ is different from $w$. Without losing generality assume $y'\neq w$. Then, by Lemma~\ref{lemma:con-edges} there exists $y_0\in [n]$ such that $xy_0$ has the same sign as $xy$ and $xy_0, x'y'$ do not intersect in the embedding of $K_n$ in the plane. Moreover, $z(x)=z(x')$. This finishes the prove of this case. 
    
\end{enumerate}

\end{enumerate} 

\end{proof}

\thmthree*

\begin{proof}

We can assume without losing generality that $z$ is a non-decreasing coloring. Let $\succ$ be any monomial order that refines the partial order specified by weights just as in the beginning of the current section. Let $\text{Bin}_{\mathcal{M}_{n,z}}:=\{x^{\gamma^+}-x^{\gamma^-}: \gamma\in {\mathcal{M}_{n,z}}\}$. By \cite[Corollary 4.4]{St}, the set of binomials $\{x^{\gamma^+}-x^{\gamma^-}:\gamma\in \ker_\Z\!\DC_{\!n,z}\}$ contains every reduced Gr\"obner basis (with respect to any monomial order) of $\IDC$. Hence, to show $\text{Bin}_{\mathcal M_{n,z}}$ is a Gr\"obner basis, it is enough to prove that the leading term of any binomial $x^{\gamma^+}-x^{\gamma^-}\in \IDC$ is divisible by a monomial $x_{uv}x_{u'v'}$ where $uv, u'v'\in E(\gamma)$ do not intersect in the embedding of $K_n$ in the plane and $\{z(u), z(v)\}\cap\{z(u'), z(v')\}\neq \emptyset$.
Assume that $f=x^{\gamma^+}-x^{\gamma^-}\in \IDC$ with $\text{in}_\succ(f)=x^{\gamma^+}$, is a minimal counterexample, in the sense it is not divisible as described above, and minimal in the sense that $f$ has minimal weight. Here the weight of a binomial is the sum of the weights of its two monomials. This means that every pair of positive edges $uv, u'v'\in E(\gamma)$ with $z(u)=z(u')$ intersect in the embedding of $K_n$ in the plane. Furthermore, we can assume that every pair of negative edges $uv, u'v'$ with $z(u)=z(u')$ intersect in the embedding of $K_n$ in the plane as well. Otherwise, we can reduce ${x}^{\gamma^-}$ modulo $\text{Bin}_{\mathcal{M}_{n,z}}$ to get a counterexample of smaller weight. On the other hand, the existence of $\gamma$ would contradict Proposition~\ref{prop:non-intersecting}. Hence no such binomial $x^{\gamma^+}-x^{\gamma^-}\in \IDC$ could exist. Therefore, $\text{Bin}_{\mathcal{M}_{n,z}}$ is a Gr\"obner basis for $\IDC$ with respect to $\succ$.

\end{proof}

\begin{rmk}\label{rmk:Grobner-basis}

The combinatorial description for the Gr\"obner basis in Theorem~\ref{thm:Grobner_basis} has a direct implication for the combinatorics of the polytope $\mathcal P_{\scriptsize{\DC}_{\!n,z}}:=\text{conv}(a_{uv}:1\leq u<v\leq n)$ defined as the convex hull of the column vectors of $a_{uv}$ of $\DC_{\!n,z}$. More specifically, as a consequence of \cite[Theorem 8.3]{St}, the Gr\"obner basis of $\IDC$ described in Theorem~\ref{thm:Grobner_basis} induces an \emph{unimodular} regular triangulation $\mathcal T_\succ$ of $\mathcal P_{\scriptsize{\DC}_{\!n,z}}$. Following ideas analogous to \cite[Remarks 2.5]{DeLoera1995hypersimplex}, this triangulation enables the computation of the Ehrhart polynomial (which, in this case, equals the Hilbert polynomial) of $\mathcal P_{\scriptsize{\DC}_{\!n,z}}$. For example, for $k=2$ and any $k$-coloring $z:[n]\to [2]$ with $n_i=|z^{-1}(i)|$, we have that the Hilbert polynomial of $\IDC$ is given by 
\[
H_{\scriptsize{\DC}_{\!n,z}}(r)=\text{card}\left(r\cdot \mathcal P_{\scriptsize{\DC}_{\!n,z}}\cap \Z^{n+3}\right)=\sum_{\tau\in \mathcal W_{r,3}} a_\tau,
\]
where $\mathcal W_{r,3}=\{\tau\in \N^{3}: \sum_{1\leq i\leq j\leq 2}\tau_{i,j}=r\}$ is the set of \emph{weak $3$-partitions} of $r$ and  
\begin{align*}
a_\tau=&\binom{n_1+2\tau_{1,1}+2\tau_{1,2}}{n_1-1}\binom{n_1+2\tau_{1,1}+2\tau_{1,2}}{n_1-1} -n_1\binom{n_1-2+\tau_{1,1}+\tau_{1,2}}{n_1-1}\binom{n_1+2\tau_{1,1}+2\tau_{1,2}}{n_1-1}\\
&-n_2\binom{n_1+2\tau_{1,1}+2\tau_{1,2}}{n_1-1}\binom{n_2-2+\tau_{2,2}}{n_2-1}+n_1n_2\binom{n_1-2+\tau_{1,1}+\tau_{1,2}}{n_1-1}\binom{n_2-2+\tau_{2,2}}{n_2-1}
\end{align*}
for every $\tau\in \mathcal W_{r,3}$. Similar formulas can be derived for $k>2$.

When the $k$-coloring $z$ is constant, $\mathcal P_{\scriptsize{\DC}_{\!n,z}}$ is linearly isomorphic to the \emph{second hypersimplex $\Delta_n(2)$}. In this specific case, the triangulation $T_\succ$ has been thoroughly described in \cite{DeLoera1995hypersimplex}. However, a general combinatorial understanding of $\mathcal P_{\scriptsize{\DC}_{\!n,z}}$ and the induced triangulation $\mathcal T_\succ$ is not yet known.

\end{rmk}

\section*{Acknowledgements} 

FAH and JDL are partially supported by NSF Grant No. DMS-1818969. FAH is grateful for the support received through NSF TRIPODS Award No. CCF-1934568. SP is partially supported by DOE award \#1010629 and the Simons Foundation Collaboration Grant for Mathematicians \#854770.
Part of this research was performed while SP and FAH were visiting the Institute for Mathematical and Statistical Innovation (IMSI), which is supported by the National Science Foundation (Grant No. DMS-1929348).

\bibliography{three-way-tables,AlgStatAndNtwks,MarkovBases}
\bibliographystyle{abbrvnat}

\end{document}